\documentclass[12pt]{article}
\usepackage[latin1]{inputenc}
\usepackage{amsfonts}
\usepackage{amsmath}
\usepackage{amssymb}
\usepackage{amsthm}
\usepackage{bigints}
\usepackage[labelfont=bf]{caption}
\usepackage{color}
\usepackage{float}
\usepackage{footmisc}
\usepackage{geometry}
\usepackage{graphicx}
\usepackage{epstopdf}
\usepackage{hyperref}
\usepackage{hypcap}
\usepackage{mathtools}
\usepackage{natbib}
\usepackage{pdfpages}
\usepackage{pgfplots}
\usepackage{rotating}
\usepackage{setspace}
\usepackage{subfig}
\usepackage{tikz}
\usetikzlibrary{calc}
\usetikzlibrary{patterns}
\usepackage{xcolor}

\pgfplotsset{compat=1.13}

\DeclareMathOperator*{\argmax}{arg\,max}
\hypersetup{
colorlinks,
linkcolor={red!50!black},
citecolor={blue!50!black},
urlcolor={blue!80!black}
}

\newcommand*{\Resize}
[2]{\resizebox{\textwidth}{!}{$#2$}}
\newcommand\myeq{\mathrel{\stackrel{\makebox[0pt]{\mbox{\normalfont\tiny
l'Hopital}}}{=}}}

\newtheorem{claim}{Claim}

\newtheorem{corollary}{Corollary}

\newtheorem{lemma}{Lemma}

\newtheorem{proposition}{Proposition}

\begin{document}

\title{Uncertain Product Availability in Search Markets\thanks{	
I thank seminar participants at the Vienna Graduate 
		School of Economics and Kelley School of Business 
		for their helpful comments, especially Jackson 
		Dorsey, Daniel Garcia, Marc Goni, Rick Harbaugh, 
		Maarten Janssen, Aaron Kolb, Eeva Mauring, Marilyn 
		Pease, Eric Rasmusen, Karl Schlag, Matan Tsur, 
		Stephanos Vlachos, and Matthijs Wildenbeest. I am 
		also grateful for helpful discussions with 
		participants of QED/Jamboree 2019, Belgrade 
		Young Economists Conference 2019, EEA-ESEM 2019, 
		EARIE 2019, 14th BiGSEM Doctoral Workshop on 
		Economic Theory, and especially Anna Obizhaeva, 
		David Ronayne, Andrew Rhodes, and Chris Wilson.  
		Financial support from uni:docs Fellowship Program 
		and the research project entitled ``Information 
		acquisition, diffusion, and provision'' from the 
		University of Vienna is acknowledged.} }%

\author{Atabek Atayev \thanks{\small Corresponding author; 
e-mail: atabek.atayev@zew.de}\\
{\small ZEW---Leibniz Centre for European Economic Research in 
Mannheim}\\
{\small L 7, 1, 68161 Mannheim, Germany}}%

\date{\today}

\maketitle

\begin{abstract}
\noindent In many markets buyers are poorly informed about which 
firms sell the product (product availability) and prices, and 
therefore have to spend time to obtain this information.  In 
contrast, sellers typically have a better idea about which 
rivals offer the product. Information asymmetry between buyers 
and sellers on product availability, rather than just prices,
has not been scrutinized in the literature on consumer search. 
We propose a theoretical model that incorporates this kind of 
information asymmetry into a simultaneous search model. Our key 
finding is that greater product availability may harm buyers by 
mitigating their willingness to search and, thus, softening 
competition. 
\newline


\end{abstract}

\sloppy
%%%%%%%%%%%%%%%%%%%%%%%%%%%%%%%%%%%%%%%%%%%%%%%%%%%%%%%%%%%%%%%%%%%%%%%%%%%%%%	

\pagebreak 
%%%%%%%%%%%%%%%%%%%%%%%%%%%%%%%%%%%%%%%%%%%%%%%%%%%%%%%%%%%%%%%%%%%%%%%%%%%%%%%%

\section{Introduction}

Initiated by \cite{stigler1961}, theory of simultaneous search 
(also known as \textit{nonsequential} search and 
\textit{fixed-sample-size} search) has been widely employed to 
study observed inefficiencies in markets, where buyers wish to 
gather information quickly and the outcome of search is observed 
with a delay (\cite{morganmanning1985}).%
\footnote{Examples include consumer financial credit markets
	(e.g., \cite{allenetal2013}, \cite{galenianosgavazza2020}), 
	labor markets (e.g.,  \cite{gautieretal2016}, 
	\cite{baggerlentz2019}), and markets for consumer retail 
	goods (e.g., \cite{moragagonzalezwildenbeest2008}, 
	\cite{chandratappata2011}, \cite{honka2014}, 
	\cite{linwindenbeest2020}, \cite{murryzhou2020}).}
The early studies in this field focus on how search costs 
mitigate buyers' incentive to discover prices (e.g., 
\cite{salopstiglitz1977}, \cite{macminn1980}, 
\cite{burdettjudd1983}).  In many markets, however, buyers 
face uncertainty not only about prices but also product 
availability. A consumer looking for a loan does not know
which banks will approve her application, a consumer planning a 
renovation does not know which companies provide the 
desired service, and a government agency soliciting bidders does 
not know which, and how many, bidders will participate in an 
auction. Later search models account for uncertain product 
availability (e.g., \cite{janssennon2009}, \cite{lester2011}), 
yet they typically assume that an individual seller, just like 
buyers, does not know whether rivals supply the product.  With 
the rise of data-gathering and -processing technologies, it has 
become evident that sellers have a good idea about their 
competitors, including whether they supply the product and are 
not capacity constrained.%
\footnote{Evidence suggests that sellers are likely to
	know which rivals supply the product.  For instance, 
	IBISWorld estimates that annual spending for acquisition 
	of competitor information by companies in the United States 
	was at $\$ \ 2$ billion in the first half of 2010s (see, 
	\cite{gilad2015}). Most large companies have employees 
	whose tasks are to gather and analyze information about 
	their rivals (see, \cite{billandetal2010}).  These examples 
	imply that, if firms invest in learning rival firms' 
	business decisions, they are likely to have a good idea	
	about product availability at their rivals.}

Despite the prevalence of information asymmetry on product 
availability in search markets, they remain understudied.  With 
this paper we aim to narrow this gap.  We derive novel results 
on the impact of product availability on market outcomes.

Our model is an extension of the canonical model of 
\cite{burdettjudd1983}.%
\footnote{\cite{burdettjudd1983}'s model of simultaneous search
	has been employed to study price competition in both 
	homogeneous goods markets (e.g., \cite{macminn1980}, 
	\cite{janssenmoraga2004}) and horizontally differentiated 
	goods markets (e.g., \cite{andersonetal1992}, 
	\cite{moragaetal2021}), competition by multiproduct firms 
	(e.g., \cite{mcafee1995}), price dynamics (e.g., 
	\cite{fershtmanfishman1992}, \cite{yangye2008}), and labor 
	markets (e.g., \cite{burdettmortensen1998}, 
	\cite{acemoglushimer2000}).  It also serves as a convenient 
	model for empirical studies employing structural estimation 
	techniques (e.g., \cite{hongshum2006}, 
	\cite{moragagonzalezwildenbeest2008}, \cite{tappata2009}, 
	\cite{delossantosetal2012}, \cite{honka2014}, 
	\cite{galenianosgavazza2020}).}
The standard features of the model are that identical 
sellers produce homogeneous goods and compete on prices, while  
buyers need to engage in costly search to discover prices.  
We extend the model by assuming that $n$ out of $N$ potential 
sellers supply the product, where $n$ is a random variable.  We 
say that the product is more available if the probability of a 
high number of sellers offering the product rises and, 
accordingly, the probability of a low number of such sellers 
falls.  Sellers know which of them offer the product and set 
prices conditional on the number of competitors.  In contrast, 
buyers do not know which, and how many, sellers supply the 
product.  Therefore, buyers cannot condition their search 
strategy on the number of actual sellers and will incur the 
search cost irrespective of whether the searched seller does or 
does not have the product.

Our main finding is that as the product becomes more available, 
buyers may be worse-off.  We identify two effects. First, there 
is a direct competition effect.  For a given buyer search 
strategy, the expected price declines with the realized number 
of sellers offering the product, i.e., with $n$.  This is simply 
because sellers compete more intensely as they face more 
competitors.  Since greater product availability means that they 
are more likely to face a larger number of competing sellers, 
buyers pay lower prices in expectation. There is, however, also 
an indirect search effect, which is anti-competitive.  With 
greater product availability, buyers' search incentive 
declines.  This happens for two reasons.  First, it is simply 
easier to find the product and compare different deals.  Second, 
the expected level of price dispersion falls.  The 
(\textit{ex-interim}) level of price dispersion has an inverse U-
shape with respect to the realized number of sellers. Then, the 
expected (or \textit{ex-ante}) level of price dispersion 
decreases if greater product availability means that a moderate 
number of sellers is less likely and a high number of sellers is 
more likely to be in the market.  If, however, prices do not 
differ much across sellers in expectation, buyers have less 
incentive to compare deals.  As buyers compare fewer deals 
by  searching less intensely, sellers' market power for each 
realization of $n$ increases.  We show that this (indirect) 
anti-competitive effect of greater product availability 
dominates the (direct) competitive effect if the search cost is 
relatively small.

To check the robustness of our main result---greater product 
availability leading to less search and softer competition---we 
consider different extensions of our model.   Specifically, we 
consider markets where buyers employ \textit{noisy search} 
(\cite{wilde1977}).  With noisy search, a searching buyer 
learns of offers of an unknown number of firms, but the more 
intensely she searches, the higher the probability is that she 
learns of more offers.  Such modeling suits procurement markets 
where a government agency announces an auction via different 
platforms (e.g., newspapers, social networking websites such as 
LinkedIn), instead of soliciting bidders individually.  The 
agency is not sure how many bidders it can attract through each 
platform, but it can reach more potential bidders if it makes 
announcements on more platforms.  We demonstrate that our main 
result holds with noisy search. In another extension of our 
model, we allow for search cost heterogeneity by introducing a 
share of buyers who observe all offers for free.  Such buyers 
can represent, for instance, consumers who use price-comparison 
websites.  We report that greater product availability mitigates 
search incentives of buyers with positive search costs and 
decreases their well-being.  Finally, we discuss information 
asymmetry on product availability in sequential search markets.  
We demonstrate that greater product availability may harm 
buyers.  However, the mechanism behind this result is different 
from that in the model of simultaneous search and in line with 
that in traditional models of consumer search with search cost 
heterogeneity (e.g., \cite{varian1980}, \cite{stahl1989}). 

Our main result has implications in real-world markets. Products 
may be more available for a variety of reasons: an improvement 
in search/matching technology, a technological shock that makes 
production and logistics more efficient, or market entry.  In 
online markets buyers rely on search engines to find the 
product. Online search platforms upgrade their algorithm 
regularly and---although we do not have direct evidence of 
this---we believe it is self-evident that with the time (namely, 
in the long run, such as a decade) search engines provide better 
matches to buyers' search requests.  Literature on operations 
management suggests that many manufacturing companies opt for 
sharing relevant information with companies along the supply 
chain to ensure timely delivery of production resources (e.g., 
\cite{fabbejahre2008}, \cite{kim2009}, 
\cite{prajogoolhager2012}).  Clearly, timely delivery of 
production resources allows firms to relax capacity 
constraints.  Regarding policy interventions, any policy that 
triggers entry makes the product more available to 
buyers.%
\footnote{The impact of entry on market outcomes has been 
extensively studied in the literature on consumer search.  
Whereas the existing literature identifies sellers' pricing 
behavior as a leading mechanism behind the results, we show 
that buyers' search behavior is a key determinant of our 
results.  Also the results of the existing studies crucially 
depend on search cost heterogeneity, whereas our results hold 
irrespective of this assumption.  We provide a detailed 
discussion of these points in the next section.}

The rest of the paper is organized as follows.  We discuss our 
paper's contribution to the literature in the following section. 
In Section \ref{s:model} we present the model.  We provide an 
equilibrium analysis in Section \ref{s:BNE} and report our
comparative statics results in Section \ref{s:cs}. In Section
\ref{s:extensions} we analyze different extensions of our main 
model. The final section concludes.

\section{Related Literature} \label{s:literature}

Our paper contributes to several strands of the literature. One 
is the consumer search literature with uncertain product 
availability and, within this field, the studies by 
\cite{janssennon2009} and \cite{lester2011} are the closest to 
our paper.  The main difference between these papers and ours is 
that they do not consider asymmetric information: an individual 
seller, just like buyers, does not observe which other sellers 
offer the product. Therefore, sellers cannot condition their 
prices on the total number of sellers in the market. For 
instance, if there happens to be a single seller in the market, 
the monopolist simply does not know this fact and, in 
equilibrium, does not necessarily set the monopoly price.  There 
are also other important differences. Specifically, 
\cite{janssennon2009} restrict their attention to two potential 
sellers, which makes the impact of  greater product availability 
on competition straightforward. Precisely, the more available 
the product (or the more probable that there are two sellers 
rather than a monopolist), the stronger the competition.  

\cite{lester2011} examines a model with exogenously given shares 
of consumers who observe a single price and those who compare 
multiple prices in markets where firms face limited production 
capacity.  In the model, consumers do not really engage in 
search.  Instead, they choose a seller from which to make a 
purchase, realizing that any seller may run out of stock if many 
buyers choose to buy from that seller. Each consumer who 
observes only one seller's price queues only at that seller. 
Given sellers' capacity constraints, price-comparing consumers 
``compete'' to buy lower-priced products. Therefore, an 
(exogenous) increase in the share of price-comparing consumers 
spurs competition among these consumers for low-priced products, 
i.e., low-priced products become less available.  
Price-comparing buyers understand that their chance of making a 
purchase at a low price decreases and thus are ready to accept 
higher prices.  This in turn increases firms' incentive to price 
the product high.  Although a higher share of price-comparing 
consumers implies that firms compete more intensely to attract 
them, the competitive impact may be dominated by the above 
anti-competitive effect.   Other papers which study uncertain 
product availability in consumer search markets, but without the 
information asymmetry, include \cite{janssenrasmusen2002}, 
\cite{rhodes2011}, \cite{gomis-porquerasetal2017} and 
\cite{moragawatanabe2020}.   

There is a large body of literature that studies search 
frictions and uncertainty about product availability in labor 
markets.  In these studies, uncertainty is related to 
availability of a job vacancy. The models employed in these 
studies are in the spirit of \cite{lester2011}, where 
job-seekers choose to which companies to apply for a job after 
observing their offers.  However, these studies do not consider 
information asymmetry on job availability.  We refer to 
\cite{wrightetal2017} for an excellent review of the 
literature.

There is another strand of the consumer search literature 
that studies information asymmetries between buyers and 
sellers.  Yet these studies focus on information asymmetry on 
either marginal cost of production (e.g., 
\cite{benabougertner1993}, \cite{dana1994}, \cite{tappata2009}, 
and \cite{janssenetal2011}), or product quality (e.g., 
\cite{heymckenna1981}, \cite{pesendorferwolinsky2003}, 
\cite{wolinsky2005}, \cite{fishmanlevy2015}).

In a broader sense associating entry of firms with greater 
product availability, the paper also contributes to the 
literature that studies the effect of entry on competition 
(e.g., \cite{janssenmoraga2004}, \cite{chenriordan2008}, 
\cite{gabaixetal2016}, \cite{gonzalezetal2017}, 
\cite{chenzhang2018}). The papers closest to ours are ones by 
\cite{janssenmoraga2004} and \cite{gonzalezetal2017}. The 
crucial difference between these papers and ours is that, in 
these papers, both consumers and sellers know which sellers 
offer the product.  In addition, \cite{janssenmoraga2004} let a 
share of consumers have zero search cost and the remaining 
consumers have a positive search cost which results in a bimodal 
search cost distribution, and \cite{gonzalezetal2017} allow for 
a continuous distribution of search costs.  
\cite{janssenmoraga2004} report that, if search costs are small 
(for buyers with a positive search cost) and the number of 
sellers is sufficiently high, the expected price rises with each 
additional entry.  The reason for this is that competition for 
consumers with zero search costs increases with the number of 
firms.  As the number of firms reaches a tipping point, firms 
find it optimal to ripoff buyers with a positive search cost, 
instead of competing intensely for consumers with zero search 
cost (as in traditional studies of \cite{varian1980} and 
\cite{stahl1989}). Similarly, \cite{gonzalezetal2017} 
demonstrate that, if search costs are relatively dispersed and 
the number of sellers is high enough, an additional firm entry 
results in higher prices.  The reasoning behind this result is 
similar to that of \cite{janssenmoraga2004}.  

The only other paper that accounts for information asymmetry on 
product availability in search markets is one by
\cite{parakhonyaksobolev2015}.  Whereas we analyze the question 
in a more traditional Bayesian game of incomplete information, 
they assume that buyers (who search sequentially) do not have a 
prior and want to minimize \textit{regret} instead of maximizing 
utility.  The authors report that the equilibrium expected price 
paid by buyers is invariant to changes in the number of 
sellers.  Their reasoning is as follows.  Given the presence of 
consumers with zero search cost and those with a positive search 
cost, an increase in the number of sellers has the same two 
(opposing) effects on prices as those in 
\cite{janssenmoraga2004}. The interaction of these effects 
pushes up prices.  However, there is also an additional effect. 
Consumers with positive search costs search more intensely as 
the number of sellers increases and, hence, are more likely to 
find a lower price.  This effect offsets the interaction of the 
above two effects, so that the expected price paid by buyers 
does not change. 

We finally note that the intuition behind our main result 
on the detrimental impact of more product availability on 
competition is similar to those in studies by 
\cite{fershtmanfishman1994} and \cite{armstrongetal2009b}. These 
papers report that price caps raise the expected price paid by 
buyers in non-sequential search markets. The reason for this is 
that an effective price cap limits the range of prices sellers 
can set.  This reduces price dispersion, which in turn mitigates 
buyers' incentive to search. Just as price caps alleviate price 
dispersion, a higher probability of more sellers offering the 
product \textit{may} make prices less dispersed in our model 
(recall the inverse-U-shaped relationship between the 
(\textit{ex-interim}) level of price dispersion and the number 
of sellers).

\section{Model}\label{s:model}

We now present our main model.  In the first part of this 
section we simply lay out our model.  In the second part we 
argue in favor of our model's assumptions.

\subsection{Assumptions, Timing, and Equilibrium Concept}

In the model there are $N\geq 3$ potential sellers, which we 
call \textit{firms}. $N$ is assumed to be finite. Nature chooses 
$n$ number of entrants, or simply \textit{sellers}, where $0\leq 
n \leq N$. The probability with which nature chooses $n$ 
entrants is given by $\theta_n$, so that $\sum_{n=0}^{N} 
\theta_n=1$. Let $\theta$ represent the probability mass 
function (PMF) and $\Theta$ stand for the corresponding 
cumulative distribution function (CDF) so that $\Theta(n)$ is 
the probability that there are at most $n$ sellers in the 
market. Note that if $\theta_N=1$, meaning that all firms are 
active sellers, the model collapses to the traditional model of 
simultaneous search as in \cite{burdettjudd1983}, yet with 
a finite number of sellers.  Each firm observes who has entered 
the market. Sellers produce homogeneous goods at a marginal cost 
normalized to zero and compete on prices. Since mixed strategies 
are allowed, we let $x_{nj}(p)$ be the probability that seller 
$j$ charges a price greater than $p$ when there are $n$ number 
of sellers in the market.

The demand side of the market is represented by a unit mass of 
buyers, or \textit{consumers}.  Each consumer has an inelastic 
demand for a unit of a product, which she values at $v>0$.%
\footnote{We can think of $v$ as the effective reservation price 
	of consumers. Specifically, if we let $r$ be the actual 
	reservation price and $w>0$ be the outside option of each 
	consumer, then $v = r-w$. Clearly, for $w\geq r$, buyers 
	never participate in the market. The paper focuses on the 
	interesting case of $r>w$.}
\textit{Ex-ante}, consumers do not know which (if any) firms are 
active sellers or what the sellers' prices are. In order 
to buy a product, a consumer has to engage in costly search and 
learn of at least one price. Search is simultaneous, meaning a 
consumer requests prices from $k$ number of firms 
simultaneously, after which the search is terminated. We let 
$c>0$ denote the search cost. Following the majority of 
literature on consumer search, we assume that searching one firm 
is free.%
\footnote{
	This assumption does not affect the main results 
	qualitatively.  If we allow each search to be costly, then 
	in the trivial equilibrium stated in Proposition 
	\ref{prop:diamond} buyers do not search and, thus, there is 
	no trade.}
Finally, let $q_{k}$ stand for the probability that buyers 
search $k$ firms so that $q$ represents the search probability 
distribution.  An alternative explanation is that $q_k$ is the 
share of consumers who search $k$ firms.  As searching one firm 
is weakly dominant than not searching at all, we set $q_{0}=0$ 
for the rest of the paper.

The timing of the game is as follows. First, nature chooses a 
number of sellers that enter the market. Each firm observes 
whether itself or any other firms entered the market. 
Consumers do not have this information.  Second, sellers 
simultaneously set prices. Third, without knowing prices, 
consumers choose the number of firms to visit. Consumers who 
observe at least one price may make a purchase.

The solution concept is Bayesian-Nash equilibrium (BNE).  Let 
$x_{-jn}(p)$ represent the equilibrium pricing strategies of all 
sellers, but seller $j$, in a market with $n$ sellers.  Also let 
$\Pi_{nj}(p,x_{-jn}(p))$ denote the expected profit of seller 
$j$ that charges $p$ given pricing strategies of other sellers. 
Then, letting $\overline{\Pi}_{nj}\geq 0$ be some constant for 
each $n$ and $j$, we define a BNE as a collection of price 
distributions $\left(x_{n1},...,x_{nn}\right)_{n=1}^N$ and 
search probability distribution $q$ such that for each $n$ (a) 
$\Pi_{nj}(p,x_{-jn}(p)) \geq \overline{\Pi}_{nj}$ for all $p$ in 
the support of $x_{nj}(p)$, $\forall j$, and (b) 
$\Pi_{nj}(p,x_{-jn}(p)) \leq \overline{\Pi}_{nj}$ for all $p$, 
$\forall j$; (c) each consumer searching $k$ firms obtains no 
lower utility by searching any other number of firms for all 
$q_k>0$ and (d) $\sum_{k=1}^N q_k =1$.

Next, it is useful to note that the probability that a consumer 
observes $m$ prices, when searching $k$ firms in a market with 
$n$ sellers, follows a hypergeometric PMF. This probability, 
denoted by $\alpha_{nk,m}$, is 
\begin{equation*}
\alpha_{nk,m} = 
\frac{\binom{N-n}{k-m}\binom{n}{m}}{\binom{N}{k}},
\end{equation*}
where for any two positive integers $I_1<I_2$, we let 
$\binom{I_1}{I_2} = 0$.  We also define 
\begin{equation*}
\alpha_{nk}(x) \equiv \sum_{m=0}^n \alpha_{nk,m}x(p)^m
\end{equation*}
to be the probability generating function, where $x(p)^m\equiv 
[x(p)]^m$.  In the appendix we provide some properties of the 
hypergeometric PMF which we will employ in our analysis.

\subsection{Discussion of Assumptions}\label{ss:model_discussion}

We now discuss the following two assumptions of the model: 
goods are homogeneous and search is simultaneous.  Regarding the 
first assumption, there are numerous real-world markets where 
products are fairly homogeneous across sellers.  In financial 
credit markets, for instance, the main difference across lenders 
is the interest rate, and this is certainly a homogeneous good.  
In procurement markets, certain construction materials, such as 
ready-mixed concrete and plywood, or renovation services, such 
as painting walls and installing electric outlets, are fairly 
homogeneous across sellers.  

Regarding our second assumption, it has been well established 
that simultaneous search is optimal in markets where a buyer 
wishes to gather information quickly and it takes time for firms 
to quote prices.  \cite{morganmanning1985} demonstrate that this 
is so even if buyers learn through search not only of prices but 
also of market-fundamentals, such as product availability.   To 
illustrate this argument, we consider a market where a 
government agency wishes to procure renovation services.  
Suppose that on average it takes a week for a renovation company 
to reply to the agency's request, e.g., to inspect the office 
rooms and quote a price.  If the agency searches sequentially, 
it first contacts one of the companies, waits around a week to 
receive a reply from it, and only then decides whether to 
contact another company.  Although sequential search allows the 
agency to better evaluate product availability with each search 
round, it is time-costly.  In our example, it takes a month (on 
average) to contact four companies.  However, the agency could 
contact all four companies on the same day and hence expect 
their replies within a week.  

The time-efficiency of simultaneous search is especially 
relevant if the search has to be terminated at some point owing 
to some exogenous deadline.%
\footnote{Implying a sharp rise in the search cost, deadlines
	are also consistent with the argument that the search cost 
	is convex in the number of searches in sequential search 
	markets (e.g., \cite{ellisonwolitzky2012}, 
	\cite{carlinederer2019}).  It can be shown that under 
	certain conditions, predictions of sequential search models 
	with convex search coincide with those of simultaneous 
	search models with linear search costs.}
Specifically we are referring to deadlines which may occur 
naturally or be a consequence of regulations.  In a market for 
home renovation services, a consumer who starts searching at the 
end of autumn may set a tight deadline to find a service 
provider because she does not wish to undertake a renovation in 
cold weather.  If, within the deadline, the consumer does not 
find a company that offers the desired service, she may prefer 
delaying the renovation until the following year.  Deadlines in 
public procurement markets are set by law.%
\footnote{\cite{cursor2019} reports that more than
	a quarter of the public procurement of medicines in the 
	Russian Federation in $2019$ was declared unsuccessful 
	because no bid was submitted within the announced deadlines.}

Finally, empirical studies confirm that simultaneous search is 
prevalent in certain markets.  For instance, 
\cite{delossantosetal2012} and \cite{honkachintagunta2017} 
demonstrate that, in online markets for books and car insurance 
respectively, models of simultaneous search predict buyers' 
search behavior better than those of sequential search.

\section{Equilibrium Analysis}\label{s:BNE}

We start our analysis by identifying consumers' search 
strategies which can be a part of a BNE.  Subsection 
\ref{ss:obs} serves this purpose.  Here we first demonstrate 
the existence of a trivial equilibrium where consumers do not 
search more than one firm.  As the existence of this equilibrium 
is fragile to model assumptions (as we show below) and the 
predictions of this equilibrium are not realistic, we 
focus on BNEs where some consumer search more than one firm, 
i.e., $q_1<1$.  We call such search behavior an \textit{active} 
search.  We show that in any BNE with active search, buyers 
either search $k$ firms where $2\leq k\leq N-1$ or randomize 
over searching $k$ and $k+1$ firms where $1\leq k\leq N-1$.

We then proceed to construct those two types of BNEs with active 
search.   To do that, we employ the following steps.  In 
Subsection \ref{ss:kk+1}, we assume that consumers randomize 
between searching $k$ and $k+1$ firms, and find the optimal 
pricing strategies of sellers.  Given these pricing strategies, 
we check whether consumers indeed find it optimal to randomize 
over searching $k$ and $k+1$ firms.  In Subsection \ref{ss:k}, 
we apply the same steps, but we consider a case in which all 
buyers search $k$ firms.

We demonstrate that a BNE with active search definitely exists 
if the search cost is not too high.  Generically, there is a
multiplicity of equilibria.  We focus on stable equilibria and 
moreover establish, in Subsection \ref{ss:stability}, that a 
locally stable BNE with active search is unique for sufficiently 
small search costs.

\subsection{Preliminary Results}

\label{ss:obs}

Our first result is that an equilibrium, where consumers search 
no more than one firm, always exists.  If buyers do not search 
more than one firm, they do not compare prices. It is then 
optimal for sellers to charge the monopoly price $v$. Such 
pricing justifies the above search strategy of buyers, as buyers 
receive zero payoff whether or not they purchase a product, yet 
searching more than one firm is costly. This is a well-known 
result in models of both sequential (\cite{diamond1971}) and 
simultaneous search (\cite{burdettjudd1983}), and is known as 
the \textit{Diamond parardox}.  

\begin{proposition} \label{prop:diamond} For any $c>0$, there 
exists an equilibrium where sellers set the monopoly price $v$ 
and consumers search no more than one firm: $q_1=1$.
\end{proposition}

It is well-known that the existence of a Diamond-paradox type of 
equilibrium is fragile to small changes in the model 
assumptions.  The equilibrium ceases to exist if, for instance, 
we introduce a very small share of consumers who learn of 
multiple offers during a single search (\cite{wilde1977}) or who 
have zero search cost (e.g., \cite{salopstiglitz1977}, 
\cite{stahl1989}).%
	\footnote{Many studies show cases of how in equilibrium 
	some of the ex-ante identical consumers observe multiple 
	prices (\cite{butters1977}, \cite{robertstahl1993}, 
	\cite{atayevjanssen2019}).}
In contrast, equilibria where some consumers search more than 
one firm are robust to such assumptions, as we show in 
Section \ref{s:extensions}. 

Therefore we turn our attention to equilibria where some 
consumers search at least two firms.  Our next two results limit 
search strategies of buyers which can be a part of an 
equilibrium with active search. To state the results, we let 
$\underline{n} \geq 2$ represent the lowest number of oligopoly 
sellers that are drawn into the market with a strictly positive 
probability. This implies that 
$\theta_2=...=\theta_{\underline{n} -1}=0$ while 
$\theta_{\underline{n}}>0$.

\begin{lemma}
\label{lem:qN=1} For any $c>0$ and $\underline{n}\geq 2$, in a 
BNE it cannot be that $\sum_{k=N-\underline{n}+2}^{N}q_k=1$.
\end{lemma}

The reasoning is by contradiction. Assume that, if there are 
at least $\underline{n}$ sellers in a market, consumers see 
two prices for certain, i.e., $\sum_{k=N-\underline{n}+2}^{N} 
q_k = 1$. Then, sellers optimally price as follows. The 
monopolist sets a price equal to $v$.  Sellers price the product 
at the marginal cost of production if there are at least 
$\underline{n}$ number of sellers in the market.  This argument 
stems from the observation that, in these markets, the price of 
an individual seller is always compared with at least one other 
price.  Therefore, an individual seller does not want to be the 
highest-priced one and, in the case of a tie in prices, 
undercutting is profitable.  Given the above pricing strategies 
of sellers, the consumer---who searches so as to make sure she 
observes at least two prices when the number of sellers in the 
market is at least $\underline{n}$---either faces a market with 
at most one seller and receives zero surplus (irrespective of 
whether or not she makes a purchase) or faces a market with at 
least $\underline{n}$ sellers and certainly makes a purchase, 
receiving a surplus equal to $v$ from the purchase. However, she 
can search one firm less, receive the same surplus, and save on 
search cost. Thus we arrive at a contradiction.

An important implication of Lemma \ref{lem:qN=1} is that, in any 
equilibrium with active search, sellers play mixed-strategy 
pricing in the market with $\underline{n}$ number of sellers. 
This is because some consumers search at least two firms and 
observe two prices, and consumers do not search so intensely 
that some buyers observe one price. On the one hand, sellers 
have an incentive to ripoff the latter type of buyers---also 
known as \textit{captive} or \textit{locked-in} consumers---by 
pricing the product high. On the other hand, sellers wish to 
price the product low to attract price-comparing consumers.  
The interaction of these two opposing incentives causes price 
dispersion.  Using this fact, the following proposition narrows 
down even more the search strategies of buyers in an equilibrium 
with active search.

\begin{proposition}\label{prop:search}
	A BNE with $q_1<1$ exists if, and only if, 	
	$\theta_0+\theta_1<1$. In any such BNE, there exists $k$ 
	such that for $2\leq k \leq N-\underline{n}+1$ it must be 
	that $q_k=1$, or for $1\leq k\leq N-\underline{n}+1$ it must 
	be that $0<q_k <1$ and $q_k+q_{k+1}=1$ .
\end{proposition}

It is straightforward that if there is at most one seller in the 
market, i.e., $\theta_0+\theta_1=1$, consumers do not search 
actively.  To understand the second part of the proposition, we 
first let $X$ denote the ex-ante hypothetical price 
distribution, which we define as follows.  This distribution 
assigns the probability weight of not finding the product to 
price $v$, as the buyer receives zero surplus both when she does 
not make a purchase and when she makes a purchase at price $v$.  
For instance, if at most one seller can be in the market 
(meaning that $\Theta(1)=1$), $X$ assigns all probability 
mass to $v$.  $X$ is non-degenerate as prices are dispersed in 
the market with $\underline{n}$ number of sellers, a fact 
implied by Lemma \ref{lem:qN=1}.  Then, we can denote the added 
benefit of searching, say, the $k+1$th firm as the difference 
between the expected minimum of $k$ hypothetical prices and that 
of $k+1$ hypothetical prices.  As the distribution of the 
minimum of $k$ hypothetical prices is $1-(1-X(p))^k$, the 
difference between the $k+1$th order statistic and the $k$th 
order statistic is $X(p)^k(1-X(p))$.  This is clearly decreasing 
in $k$, meaning that the expected benefit of searching one more 
firm decreases with $k$.  However, as the expected cost of 
searching one more firm is constant, it must be optimal for 
consumers to search the same number of firms or randomize over 
searching two adjacent numbers of firms. 

We now present properties of price distribution in equilibrium 
with active search. 

\begin{proposition}\label{prop:pricing} 
In any BNE with active search, the pricing strategy of sellers 
is symmetric and unique for each $n$.  Furthermore, for each 
$n$ the equilibrium price distribution is either degenerate 
with a unit probability mass at either $v$ or $0$, or is 
non-degenerate with the highest price being $v$ and contains no 
mass points or flat regions in its support. 
\end{proposition}

The reasoning is as follows. Clearly, a monopolist seller always 
charges $v$, meaning there is a unit probability mass at the 
monopoly price.  If all consumers search at least two firms 
(note that this does not mean that all consumers observe two 
prices for $n=\underline{n}$ and thus does not violate Lemma 
\ref{lem:qN=1}), consumers learn at least two prices when all 
firms happen to be active sellers, i.e., when $n=N$.  In this 
case sellers charge a price equal to the production marginal 
cost in equilibrium, implying that there is a unit probability 
mass at $p=0$.  Notice that such pricing does not necessarily 
arise in equilibrium, e.g., there will be price dispersion for 
any $\underline{n}\leq n\leq N$ if buyers randomize between 
searching one firm and searching two firms.  Also note that, for 
any realization of $n$ where sellers play pure-strategy pricing, 
there is no asymmetry in pricing strategies.   

Any asymmetry or multiplicity of optimal pricing strategy 
may then occur only in markets with price dispersion. 
\cite{johnenronayne2020} establish that mixed-strategy pricing 
of sellers in equilibrium is symmetric and unique if the share 
of buyers who exactly observe two prices is strictly positive.  
We show in the appendix that there is a strictly positive share 
of buyers who exactly observe prices for any $n$ such that there 
is an equilibrium price dispersion. To ease the notation we drop 
seller-specific indices to simply write $x_n$ and 
$\overline{\Pi}_n$ to denote the equilibrium price distribution 
and the equilibrium profit in a market with $n$ sellers for the 
rest of the paper.  

Moreover, any equilibrium non-degenerate price distribution 
$x_n$ must be atomless because if it had an atom, undercutting 
would be beneficial owing to the strictly positive share of 
consumers who compare at least two prices. Also $x_n$ cannot 
have a flat region in the support, otherwise an individual 
seller will not be indifferent between charging the lowest price 
and charging the highest price in that flat region. Furthermore, 
the highest price in the support of $x_n$ must be equal to $v$.  
It cannot exceed $v$ since a seller charging a price higher than 
$v$ does not sell to anyone.  The highest price in the support 
cannot be less than $v$ because if it were, a seller could 
improve its profit by deviating to $v$, as its expected demand 
in both cases consists of only locked-in buyers.

The last two propositions provide us with a great deal of 
information about consumers' and sellers' strategies in 
equilibria with active search but relatively little information 
about conditions under which such equilibria may exist.  The 
following two subsections address this issue.

\subsection{Mixed Search Strategy}

\label{ss:kk+1}

We start by considering the case where consumers play mixed 
strategies.  Suppose that buyers randomize between searching $k$ 
and $k+1$ firms, where $1 \leq k\leq N-\underline{n}+1$.  What 
is the optimal pricing strategy for sellers? Obviously, the 
monopolist seller always charges $v$. For $n\geq N-k+2$, the 
sellers price at the production marginal cost as all consumers 
compare at least two prices. Finally, when there are $n$ sellers 
in the market such that $2\leq n\leq N-k+1$, they set prices 
from price distribution $x_n$.  The equilibrium strategy of 
sellers is such that an individual seller is indifferent between 
setting any price in the support of the equilibrium price 
distribution and must (weakly) prefer these prices to ones which 
are not in the support.

To derive price distribution $x_n$, we first note that a 
consumer searching $k$ firms buys from seller $j$ if she visits 
the seller and observes no lower price than the seller's price. 
Therefore, seller $j$ pricing at $p$ sells to this consumer with 
probability 
\begin{equation*}
\sum_{m=1}^{n} \frac{\binom{N-n}{k-m}\binom{n}{m}m}{\binom{N}{k}n}%
x_n(p)^{m-1} = \frac{1}{n}\sum_{m=0}^{n} \frac{\binom{N-n}{k-m}\binom{n}{m}}{%
\binom{N}{k}}mx_n(p)^{m-1} = \frac{\alpha^{\prime }_{nk}(x_n(p))}{n}.
\end{equation*}
We next let $\beta_{nk}(x) \equiv q_k\alpha_{nk}(x) 
+(1-q_k)\alpha_{nk+1}(x)$ so that $\beta_{nk,m} \equiv q_k 
\alpha_{nk,m} + (1-q_k)\alpha_{nk+1,m}$ is the total share of 
consumers who observe $m$ prices. Then, seller $j$ that sets 
price $p$ expects to earn 
\begin{equation*}
\Pi_{nj}(p,x_{-jn}(p)) = p \frac{\left(\beta_{nk,1} + 2 
\beta_{nk,2}x_n(p) +
3\beta_{nk,3}x_n(p)^2 +... \right)}{n} = \frac{\beta_{nk}^{\prime }(x_n(p))p%
}{n}.
\end{equation*}

As an individual seller is indifferent in terms of setting any 
price in the support of the equilibrium distribution function, 
it follows
\begin{equation}  \label{eq:xk+1}
p \beta_{nk}^{\prime }(x_n(p)) = v \beta_{nk}^{\prime}(x_n(v)).
\end{equation}
This equation implicitly and uniquely defines equilibrium 
$x_n(p)$ (recall that uniqueness follows from Proposition 
\ref{prop:pricing}).  For convenience, we will use the inverse 
function $p_n(x_n)$, which in equilibrium satisfies $p_n(x_n) = 
n \overline{\Pi}_n /\beta'_{nk}(x_n).$ Then, the lower bound of 
the price distribution, denoted by $\underline{p}_n$, solves 
$\underline{p}_n = p_n(1)$.

We now have to check whether consumers indeed randomize 
between searching $k$ firms and searching $k+1$ firms if sellers 
price the product as discussed above. To do so, we first note 
that as the density of the lowest of $m$ prices is $-m 
x_n(p)^{m-1} x_n^{\prime }(p),$ the expected price paid by a 
buyer searching $k$ firms in a market with $n$ sellers is 
\begin{equation*}
-\sum\limits_{m=1}^{n}\int_{\underline{p}_n}^{v}p\alpha_{nk,m} m
x_n(p)^{m-1}x_n^{\prime }(p)dp = - \int_{\underline{p}_n}^{v}p\alpha^{\prime
}_{nk}(x_n(p))x_n^{\prime }(p)dp = \int_{0}^{1}p_n(x_n)\alpha^{\prime
}_{nk}(x_n)dx_n,
\end{equation*}
where we obtained the last equality by changing variables from 
$p$ to $x_n(p)$. As not making a purchase is equivalent to 
paying price $v$, we define the expected \textit{virtual} price 
paid by a consumer who searches $k$ firms as
\begin{equation*}  \label{eq:Pk}
P_k \equiv  (\theta_0+\theta_1) v + 
\sum_{n=\underline{n}}^{N-k+1}\theta_n\left(\alpha_{nk,0}v
+ \int_{0}^{1} p_n(x_n)\alpha^{\prime }_{nk}(x_n)dx_n\right).
\end{equation*}
Here, we used the fact that pricing policies of sellers in 
markets with $n$ sellers such that $1< n < \underline{n}$ are 
irrelevant for consumers as, by the definition of 
$\underline{n}$, we have $\theta_n=0$ for all $1<n < 
\underline{n}$. Next, we can express the expected virtual price 
paid by consumers who search $k+1$ firms the same way as above 
by changing the respective indices from $k$ to $k+1$. Then, the 
incremental benefit of searching the $k+1$th firm is 
\begin{equation*}  \label{eq:Pk-Pk+1}
\begin{aligned} P_{k} - P_{k+1} &&=&&&\sum_{n=\underline{n}}^{N-k+1}\theta_n
\left((\alpha_{nk,0}-\alpha_{nk+1,0})v+
\int_{0}^{1}p_n(x_n)\left(\alpha'_{nk}(x_n) -
\alpha'_{nk+1}(x_n)\right)dx_n\right)\\ &&=&&& 
-\sum_{n=\underline{n}}^{N-k+1}\theta_n
\int_{0}^{1}p_n'(x_n)\left(\alpha_{nk}(x_n) - \alpha_{nk+1}(x_n)\right)dx_n,
\end{aligned}
\end{equation*}
where the second line follows from integration by parts and the facts that $%
\alpha_{nk}(0) =\alpha_{nk,0}$, $\alpha_{nk+1}(0) = \alpha_{nk+1,0}$, $%
p_n(0)=v$ and $\alpha_{nk}(1) = \alpha_{nk+1}(1) = 1$.  Clearly, 
in equilibrium, it must be the case that 
\begin{equation}  \label{eq:IC}
P_k - P_{k+1} =c.
\end{equation}

In the following lemma, we state properties of the expected 
benefit of searching the $k+1$th firm, and we provide the proof 
in the appendix.

\begin{lemma}\label{lem:LHS_IC}
	$P_{k} - P_{k+1}$ is positive and strictly concave in $q_k 
	\in (0,1)$ for each $k$ where $1\leq k \leq 
	N-\underline{n}+1$. 
\end{lemma}

The lemma informs us that, if there exists a solution to 
\eqref{eq:IC}, there must be either at least one or at most two 
solutions in $q_k \in (0,1)$.  This is due to the concavity of 
the expected benefit of searching the $k+1$th firm with respect 
to the share of consumers who search $k$ firms, $q_k$.  The 
lemma also tells us that a solution(s) exists for a nonempty 
interval of search costs and the lower bound of this interval is 
positive.  Let $\underline{c}_{k,k+1}$ and 
$\overline{c}_{k,k+1}$ represent the lower and upper bounds 
of that interval with $0\leq \underline{c}_{k,k+1} < 
\overline{c}_{k,k+1}$.  Then, according to the lemma, the bounds 
are determined as 
\begin{equation}\label{eq:c_k_k+1}
\begin{aligned}
	\underline{c}_{k,k+1} &=&& \min\left\{(P_k - P_{k+1}|q_k=0), 
	(P_k - P_{k+1}|q_k=1)\right\},\\
	\overline{c}_{k,k+1} &=&& \max_{q_k}\{P_k - P_{k+1}\}.
\end{aligned}
\end{equation}
This suggests that for values of the search cost in interval 
$(\underline{c}_{k,k+1}, \overline{c}_{k,k+1})$ there exist 
either one or two equilibria if buyers prefer searching either 
$k$ or $k+1$ firms to not searching (i.e., searching one firm).  
The following proposition demonstrates that this is indeed the 
case.

\begin{proposition}\label{prop:eq_k_k+1} 
	Let $c	\in (\underline{c}_{k,k+1},	\overline{c}_{k,k+1})$ 
	for any $1\leq k \leq N-\underline{n}+1$, 
	then there exist at least one and at most two BNEs where 
	buyers randomize over searching $k$ and $k+1$ firms.  Such a 
	BNE is given by $(\left(x_n\right)_{n=1}^N, q)$, where 
	the equilibrium price is $v$ for $n=1$, $0$ for $n\geq 
	N-k+2$ and $x_n$ is determined by 
	\eqref{eq:xk+1} for $\underline{n}\leq n \leq N-k+1$, and 
	buyers' search strategy is determined by \eqref{eq:IC}.  
	
	Furthermore, $\underline{c}_{k,k+1}=0$ for $k=1$ and 
	$k=N-\underline{n}+1$.
\end{proposition}

The proof is in the appendix.  The first part of the proposition 
shows the existence of search cost intervals under which 
equilibria in mixed strategies of buyers exist, but it does not 
tell us much about those intervals.  The second part of the 
proposition partially addresses this issue (we address this 
issue fully in Proposition \ref{prop:eq_k}).  Specifically, it 
implies that at least two equilibria exist for sufficiently 
small search costs: in one, buyers randomize over searching one 
and two firms, and in the other, they randomize over searching 
$N-\underline{n}+1$ and $N - \underline{n}+2$ firms.

\subsection{Pure Search Strategy}

\label{ss:k}

Next, we consider the case where all consumers search $k$ firms, 
where $2\leq k \leq N-\underline{n} + 1$.  It is easy to see 
that, if all consumers search $k$ firms, the monopolist seller 
charges $v$ and sellers price at the production marginal 
cost for $n\geq N-k+2$ in equilibrium. For an intermediate 
number of sellers, $2\leq n \leq N-k+1$, price dispersion arises 
and equilibrium price distribution $x_ n$ is determined by 
\begin{equation}  \label{eq:xk}
p \frac{\alpha^{\prime }_{nk}(x_n(p))}{n} = v \frac{\alpha_{nk,1}}{n} = 
\overline{\Pi}_n>0,
\end{equation}
where the inequality is due to $\alpha_{nk,1}>0$.

To check whether it is indeed optimal for buyers to visit $k$ 
firms, given the above pricing strategies of sellers, it 
suffices to find conditions (if there are such)
under which the following set of inequalities holds for $q_k=1$: 
\begin{equation}  \label{eq:IC_2}
\begin{aligned} 
	P_k - P_{k+1}\leq c,\\
	P_{k-1} - P_{k} \geq c. 
\end{aligned}
\end{equation}
The following proposition shows that there exists a nonempty 
interval of search costs such that the set of inequalities is 
satisfied. 

\begin{proposition}
\label{prop:eq_k} For any $v>0$, $N\geq 3$, $\underline{n}\geq 
2$, $0\leq \theta_0+\theta_1<1$ and $k$ such that $2\leq k \leq 
N-\underline{n}+1$, there exists 
$[\underline{c}_{k},\overline{c}_k] 
\subset [0,v]$ given by
\begin{equation}\label{eq:c_k}
	\begin{aligned}
		\underline{c}_k = (P_k - P_{k+1}|q_k=1),\\
		\overline{c}_k = (P_{k-1} - P_{k}|q_k=1),
	\end{aligned}
\end{equation}
such that for $c \in [\underline{c_k}, 
\overline{c}_k]$ there exists only one BNE where all buyers 
search $k$ firms.  The BNE is given by 
$(\left(x_n\right)_{n=1}^N, q)$ where $p=v$ for $n=1$, $p=0$ for 
$n\geq N-k+2$, $x_n$ is determined by \eqref{eq:xk} for $2 
\leq n \leq N-k+1$ and $q_k=1$.
\end{proposition}

The proof is in the appendix, and the intuition is similar to 
that behind Proposition \ref{prop:eq_k_k+1}.  Therefore we omit 
the discussion of the intuition and, instead, we point out the 
relationship between cutoff values of search cost in the two 
propositions. Observe that in equilibrium in the buyers' mixed 
search strategy, where they randomize over searching $k$ and 
$k+1$ firms, the share of consumers who search $k$ firms can 
converge to one only if the value of search cost approaches 
$\underline{c}_k$.  Similarly, in that equilibrium, the share of 
buyers who search $k+1$ firms can converge to one only if the 
value of search cost approaches $\overline{c}_{k+1}$. Formally, 
the former means that $\lim\limits_{q_k \uparrow 1} (P_k - 
P_{k+1}) = \underline{c}_k$ and the latter implies that 
$\lim\limits_{q_k \downarrow 0} (P_k - P_{k+1}) = 
\overline{c}_{k+1}$.  These two observations, along with the 
second part of Proposition \ref{prop:eq_k_k+1}, mean that if the 
search cost is not very high, there always exists an equilibrium 
with active search.  

\begin{figure}[ht]
	\centering
	\captionsetup{justification=centering}  
	\includegraphics[width=.6\linewidth]{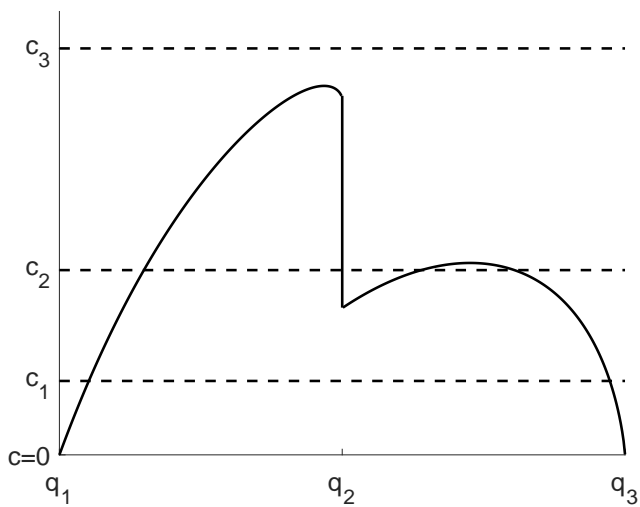}  
	\caption{Illustration of BNEs for  $N=3$, $k=2$, 
		$v=1$, $\theta_0=0$, $\theta_1=\theta_3=0.05$, and 
		$\theta_2=0.90$.}
	\label{fig:all_eq}
\end{figure}

Figure \ref{fig:all_eq} graphically illustrates the idea behind 
the discussion (here $c_1 = 0.02, c_2 = 0.05$ and $c_3 = 0.11$). 
The horizontal axis represents $q_n$ for $n \in \left\{ 1,2,3 
\right\}.$ At each of the three points on the axis, 
i.e., at each $q_n$, we have $q_n=1$.  When we move to the left 
or right of that point along the axis, $q_n$ decreases 
and $q_{n-1}$ or $q_{n+1}$, respectively, increases. For 
example, start with point $q_2$ on the horizontal axis, which 
means that $q_2=1$ and $q_1=q_3=0$. If we gradually move to the 
left along the axis, $q_2$ begins decreasing while $q_1$ begins 
increasing so that $q_1+q_2=1$ and $q_3=0$. The vertical axis of 
the graph represents values of the search cost and the expected 
benefit of searching one more firm. The solid curve stands for a 
function of the incremental benefit of searching one additional 
firm. Observe that any point on the solid line that corresponds 
to a point between $q_k$ and $q_{k+1}$ on the horizontal axis 
represents the incremental benefit of searching the $k+1$th 
firm. The dashed lines stand for different levels of search 
cost. Each intersection of the solid curve and a dashed line 
represents an equilibrium for that particular value of search 
cost. Importantly, we note the following two points. The solid 
line is continuous over $q_n$s with $\lim_{q_2\uparrow 1}(P_1 - 
P_2) = \overline{c}_2 $ and $\lim_{q_2 \uparrow 1}(P_2 - P_3) = 
\underline{c}_2$.  Moreover, the expected benefit of searching 
one more firm decreases as all buyers tend to search either one 
firm or three firms, i.e., $\lim_{q_1\uparrow 1}(P_1 - 
P_2) = \underline{c}_{1,2}=0$ and $\lim_{q_3 \uparrow 1}(P_2 - 
P_3) = \underline{c}_{2,3}=0$.  These points imply that an 
equilibrium with active search definitely exists if the search 
cost is not too high, e.g., if $c=c_1 $ or $c=c_2$.  For 
instance, there is no BNE with active search for a value of 
search cost equal to $c_3$.

\subsection{Stability}\label{ss:stability}

Subsections \ref{ss:kk+1} and \ref{ss:k} demonstrate a 
multiplicity of equilibria with active search.  For instance, 
observe in Figure \ref{fig:all_eq} that there are four BNEs with 
active search for a value of search cost equal to $c_2$.  In 
this subsection we focus on locally stable equilibria, as we 
wish to undertake comparative static analysis later.  We show 
that for sufficiently small search costs there exists a unique 
locally stable equilibrium with active search.  This equilibrium 
is characterized by buyers' mixed search strategy.

We employ a notion of stability widely applied in consumer 
search literature (e.g., \cite{burdettjudd1983}, 
\cite{fershtmanfishman1992}, \cite{janssenmoraga2004}, 
\cite{atayevjanssen2019}).  Specifically, we say that a BNE is 
locally stable if a small perturbation in search intensity of 
consumers around an equilibrium one leads to the convergence of 
the search intensity to the equilibrium one.  We realize that 
this notion of stability focuses on out-of-equilibrium behavior 
of buyers only.  In principle, one could take a different route: 
fix the search behavior of consumers and examine price 
adjustments to a small perturbation. Yet this notion of 
stability is difficult to conceive, as the equilibrium pricing 
is in mixed strategies over a compact support for certain 
realizations of $n$ and convergence to such a mixed strategy is 
difficult to conceptualize.

Having formalized the notion of stability, we now state 
conditions under which there is a unique stable BNE with active 
search in the following corollary.  The corollary is implied by 
propositions \ref{prop:eq_k_k+1} and \ref{prop:eq_k}, and its 
proof is in the appendix.

\begin{corollary}
	\label{cor:unique_BNE} For $c \in (0,\min\{\underline{c}_2, 
	\underline{c}_{N-\underline{n}+1}\})$, there exists a unique 
	locally stable BNE where consumers randomize over searching 
	$N-\underline{n}+1$ and $N-\underline{n}+2$ firms. 
\end{corollary}

The intuition is as follows. In the appendix we first show 
that only two equilibria exist for sufficiently small 
search costs.   Recall from the second part of Proposition 
\ref{prop:eq_k_k+1} that, in one of the two equilibria, buyers 
randomize between searching one firm and searching two firms 
and, in the other equilibrium, they randomize over searching 
$N-\underline{n}+1$ and $N-\underline{n}+2$ firms. To show that 
these two are the only possible equilibria for sufficiently 
small search costs, we demonstrate that lower bounds of search 
cost intervals under which other than the above two equilibria 
exist are strictly positive.  To see this point in Figure 
\ref{fig:all_eq}, note that the lower bound of the search cost 
interval where all buyers search two firms (which is represented 
by the lower point on a vertical segment of the solid curve) is 
strictly positive.  

We next show in the appendix that of the above two equilibria, 
only an equilibrium where buyers randomize over searching 
$N-\underline{n}+1$ and $N-\underline{n}+2$ firms is stable.  
To do so, we first note that if the share of consumers who 
search one firm converges to one, the expected benefit of 
searching a second firm vanishes.  This is intuitive because if 
all consumers search one firm, sellers optimally set the 
monopoly price and price dispersion vanishes.  Similarly, if the 
share of consumers who search $N-\underline{n}+2$ firms 
approaches one---meaning that they all observe at least two 
prices for certain when there are $\underline{n}$ or more 
sellers---price dispersion vanishes and an individual buyer is 
better-off searching one firm less (recall Lemma 
\ref{lem:qN=1}).  In Figure \ref{fig:all_eq}, the former point 
is illustrated by the value of the solid curve at point $q_1$ 
and the latter by the value of the curve at point $q_3$.  

Second, we know from Lemma \ref{lem:LHS_IC} that the expected 
benefit of searching $k+1$th firm in an equilibrium where buyers 
randomize over searching $k$ and $k+1$ firms is positive and 
concave in the share of buyers who search $k$ firms. This, along 
with the above two limiting results, implies that the expected 
benefit of searching a second firm is increasing in the share of 
buyers who search two firms in an equilibrium where buyers 
randomize between searching one firm and searching two firms, 
whereas the expected benefit of searching $N-\underline{n}+2$ 
firms is decreasing in the share of buyers who search 
$N-\underline{n}+2$ firms in an equilibrium where buyers 
randomize between searching $N-\underline{n}+1$ firms and 
searching $N-\underline{n}+2$ firms.  Then, in the former 
equilibrium, if the actual share of consumers who search two 
firms is smaller (larger) than the equilibrium one, the expected 
benefit of searching a second firm is lower (higher) than the 
cost of doing so.  Therefore, buyers have even less (more) 
incentive to search two firms, and buyers' search intensity 
diverges from the equilibrium one.  In Figure \ref{fig:all_eq}, 
this equilibrium corresponds to the left-most intersection of 
the dashed line $c_1$ and the solid curve.  We can apply a 
similar argument to see that an equilibrium where buyers 
randomize searching over $N-\underline{n}+1$ and 
$N-\underline{n}+2$ firms is stable.  In Figure 
\ref{fig:all_eq}, this is illustrated by the right-most 
intersection of the dashed line $c_1$ and the solid curve.

\section{Comparative Statics}

\label{s:cs}

In this section we examine the impact of changes in our two 
exogenous parameters on equilibrium market outcomes.  The 
exogenous parameters of interest are $\theta$ and $c$, which 
represent product availability and search cost, respectively.  
The market outcomes of interest are expected price 
paid by consumers and their surplus.  From now on, we will use 
the phrase \textit{expected price paid} to refer to the expected 
price paid by consumers conditional on making a purchase. (In a 
BNE in mixed strategies of consumers, the expected price paid 
would mean the \textit{average} expected price paid by consumers 
conditional on making a purchase when they randomize over 
searching two numbers of firms.)  Notice that the expected price 
paid is not the same as the virtual price, as the virtual price 
assumes that a consumer who does not learn of any price and thus 
does not make a purchase pays price $v$.   We focus on stable 
BNEs. 

A change in $\theta$ can be a product of technological 
growth or government intervention, as we discussed in the 
introduction.  Intuition tells us that greater product 
availability should benefit buyers, as they are more likely to 
find the desired product and, importantly, a greater number of 
sellers is usually associated with more intense competition. We 
demonstrate that this does not necessarily have to be the case. 
Specifically, greater product availability may not have any 
impact on market outcomes or may even harm buyers. The latter 
result is due to the detrimental effect of greater product 
availability on consumers' willingness to search.

Changes in $c$ may represent technological advances, such as 
shops creating their websites so that consumers can find out 
whether a shop has a product by means of several clicks 
instead of visiting a brick-and-mortar store. Intuitively---and 
this turns out to be the case---a smaller search cost 
strengthens buyers' willingness to search, which in turn 
increases consumers' chances of comparing prices and thus 
triggers competition.

Following the order of our analysis in Section \ref{s:BNE}, we 
first undertake a comparative static analysis in BNEs where 
buyers use a mixed strategy.  In Subsection \ref{ss:cs_k} we 
examine markets where consumers play a pure strategy.

\subsection{Consumers Playing a Mixed Strategy}

\label{ss:cs_kk+1}

We focus on a BNE that results when the search costs are 
sufficiently small. There are two reasons for this. One is that, 
following Corollary \ref{cor:unique_BNE}, we can see that there 
is 
a unique locally stable BNE with active search for small search 
costs. The other reason is that markets that have been mentioned 
in the introduction are generally characterized by low search 
costs relative to the value of the product.

We notice that, as $\sum_{n=0}^N\theta_n=1$, an increase in the 
probability that there are $i$ sellers (or $\theta_i$) must 
be accompanied by a decrease in at least one other probability, 
e.g., $\theta_j$ where $j\neq i$.  In other words, there 
are numerous ways of considering a change in the probability 
that there are $i$ sellers. To understand the main mechanism 
through which a change in the probability of $i$ sellers being 
in the market affects market outcomes, it is sufficient to focus 
on a change in $\theta_i$ that is associated with an opposite 
change in only a single $\theta_j$. In this case, it only 
matters whether $2\leq i,j\leq N-k+1$ or not, for $i\neq j$.  
Then, we need to consider only three cases: 
neither $i$ nor $j$ is in the set of integers in interval 
$[2,N-k+1]$, only $i$ or $j$ is in that set, and both $i$ and 
$j$ are in that set. We assume that $i>j$ such that an increase 
in $\theta_i$ with the associated equal decrease in $\theta_j$ 
implies that a product is more available.

In the following proposition, we state the main result of the 
section.   Specifically, we identify sufficient conditions under 
which greater product availability has a detrimental effect on 
buyers' well-being.

\begin{proposition}
\label{prop:cs_t_mix} In a stable BNE where consumers randomize 
between searching $N-\underline{n}+1$ and $N-\underline{n}+2$ 
firms for $2\leq \underline{n}\leq N-1$, an increase in 
$\theta_i$ with a corresponding decrease in $\theta_j$ 
where $i>j$

\begin{itemize}
\item[(i)] causes weakly more search, decreases the expected 
price paid, and improves consumers' well-being for $j=1$, 

\item[(ii)] does not affect the search behavior of buyers, the 
expected price paid or their well-being for $j\geq 
\underline{n}+1$, 

\item[(iii)] causes less search, raises the expected price paid 
and harms consumers' well-being for $j=\underline{n}$.
\end{itemize}
\end{proposition}

The reasoning behind (i) is straightforward. Note that, in 
equilibrium, prices are dispersed only in a market with 
$\underline{n}$ sellers and the price in a market with at least 
$\underline{n}+1$ sellers is equal to the production marginal 
cost.   As the expected prices in markets for any realization of 
$n\geq 2$ are lower than the monopoly price, the direct effect 
of a decrease in the probability that there is a monopolist 
seller is a fall in the expected price paid. If this probability 
decreases at the expense of the probability of $\underline{n}$ 
sellers being in the market, there is an additional indirect 
effect on the expected price paid. Following such a change in 
product availability, consumers search more intensely as they 
are more likely to face a market with price dispersion. 
Therefore, sellers in a market with $\underline{n}$ sellers lose 
their market power because the share of price-comparing 
consumers rises. In addition, this change in product 
availability makes it easier to find the product.  Despite the 
fact that more searching also leads to more resources spent 
on search, consumers are better-off.

The understanding behind (ii) is intuitive. Since the 
equilibrium price in a market with at least $\underline{n}+1$ 
sellers is equal to the production marginal cost, a decrease in 
the probability that there are at least $\underline{n}+1$ 
sellers, accompanied by an increase in the probability that 
there are even more sellers, does not change the expected price 
paid. Also, it does not affect consumers' search behavior, as 
this change in product availability does not affect the 
expected level of price dispersion. Finally, it does not make it 
more, or less, likely to find the product, as in markets with at 
least $\underline{n}+1$ sellers all consumers observe price 
offers of at least two sellers. Then, sellers' market power does 
not change for any realization of $n$, nor does consumers' 
well-being.

Finally, the intuition behind (iii) is as follows. First, there 
is a direct effect on the expected price paid. To show that, we 
keep consumers' and sellers' strategies fixed.  Since 
prices in a market with $\underline{n}$ sellers are bounded 
above the production marginal cost and the equilibrium price in 
a market with more than $\underline{n}$ sellers is equal to 
zero, the expected price paid decreases.  There is also an 
indirect effect. This effect takes into account how buyers and 
sellers respond to the change in product availability.  
Precisely, consumers search less following a decrease in the 
probability of $\underline{n}$ sellers being in the market 
because consumers are less likely to face a market with price 
dispersion. We know that less search raises the market power of 
sellers in a market with $\underline{n}$ sellers.  Although 
consumers save on search costs, we demonstrate in the appendix 
that the negative effect of greater product availability on 
buyers' well-being dominates its positive effects.

To illustrate this idea, we present an example with $N=3$, 
$c=0.05v$, and $\theta_2+\theta_3=1$, where $\theta_3$ increases 
from $0.1$ to $0.25$. In a unique stable equilibrium with active 
search, buyers randomize between searching two and three firms. 
This means that the prices are dispersed in a duopoly market and 
triopoly sellers charge a price equal to the marginal cost of 
production. Following the increase in $\theta_3$, the share of 
consumers who search all three firms drops from approximately 
$78\%$ to $61\%$, which is around a $22\%$ decrease. The 
expected price paid increases by around $54\%$---from $0.153v$ 
to $0.235v$. As a result, the consumer surplus (incorporating 
search costs) falls from approximately $0.758v$ to $0.685v$, 
which is a decrease of $9.6\%$.    

We next discuss an extension of our analysis of the impact of 
greater product availability on market outcomes by putting more 
structure on the change of PMF $\theta$.  In assessing how any 
such changes in product availability affect market outcomes, we 
need to check which parts of Proposition \ref{prop:cs_t_mix} 
come into play.  If different forces in different parts of the 
proposition affect the expected price paid and consumer 
well-being the same way (or do not affect market outcomes 
in different directions, such as parts (i) and (ii)), we can 
obtain clear-cut results.  For instance, suppose that a market 
can be only an oligopoly, i.e.,  $\Theta(1)=0$.  This happens 
if, say, the entry costs of at least $\underline{n}$ firms are 
sufficiently small that they always enter the market and those 
of the remaining firms are high so that they randomize between 
entering the market and not entering it.  Now assume that the
product becomes more available such that it causes first-order 
stochastic dominance (FOSD) of CDF $\Theta$.  Then, it must be 
accompanied by a decrease in the probability that there are 
$\underline{n}$ sellers in the market and an increase in the sum 
of probabilities that there are more than $\underline{n}$ 
sellers in the market.  Note that economic forces in parts (ii) 
and (iii) of Proposition \ref{prop:cs_t_mix} are relevant for 
this case, and thus consumers pay higher prices in expectation 
and are worse-off. 

However, if a similar change in product availability takes place 
in a market where a seller can be a monopolist with a strictly 
positive probability (i.e., $\Theta(1)>0$), it is ambiguous how 
market outcomes will react to this change in product 
availability.  Buyers may face a monopolist seller if all firms 
have the same entry cost, so in equilibrium they all 
randomize between entering the market and not entering it.  
Then, the above change in product availability may reduce the 
probability that there is at most one monopolist seller and that 
there are $\underline{n}$ sellers in the market, but increase 
the sum of probabilities of more than $\underline{n}$ sellers 
being in the market.  Note that in this case, economic forces in 
all three parts of Proposition \ref{prop:cs_t_mix} come into 
play.  Which of these forces prevail depends on the way one 
models a change in product availability (which at the same time 
results in the FOSD of the CDF $\Theta$).  To offer some 
insights into such situations, we provide a numerical assessment 
of the impact of greater product availability for some commonly 
used probability distributions in the online appendix.  There we 
use two versions of each probability distribution.  One of these 
versions includes a strictly positive probability that there is 
at most one seller, and the other excludes this possibility.  
For each of the probability distributions, consumers are 
better-off in the former case and worse-off in the latter case.

Now we examine the impact of a change in $c$ on market outcomes.

\begin{proposition}
	\label{prop:cs_ck+1} In any stable BNE characterized by 
	consumers randomizing over searching $k$ and $k+1$ firms, an 
	increase in $c$ causes less search and impairs consumers' 
	well-being.
\end{proposition}

It is fairly straightforward that an increase in search cost 
mitigates consumers' willingness to search. We know that less 
search is associated with greater market power of sellers. 
We also know consumers are less likely to find the product if 
they search less. Consumers still spend less resources on search 
costs. In the appendix we show that the former two negative 
effects of an increase in search cost on consumers' well-being 
dominate the latter positive effect.

\subsection{Consumers Playing a Pure Strategy}

\label{ss:cs_k}

We continue our comparative static analysis to examine BNEs 
where consumers play a pure strategy. Notice that there is a 
continuum of search costs under which such an equilibrium 
exists. This means that marginal changes in $\theta$ or $c$ do 
not affect the search behavior of buyers or, therefore, pricing 
strategies of sellers. Nevertheless, these changes affect 
consumers' well-being as well as the total expected price paid 
by consumers, as we show in the following proposition.

\begin{proposition}
\label{prop:cs_thk} In any stable BNE characterized by all 
consumers searching $k$ firms,

\begin{itemize}
\item[(i)] an increase in $\theta_i$ with a corresponding decrease in $%
\theta_j$ does not affect consumers' search behavior and

\begin{itemize}
\item[(a)] pushes down the average expected price paid and 
improves buyers' well-being for $j\leq N-k+1$,

\item[(b)] has no impact on the expected price paid or on 
buyers' well-being for $j\geq N-k+2$;
\end{itemize}

\item[(ii)] an increase in $c$ does not affect consumers' search 
behavior or the expected price paid and impairs their well-being.
\end{itemize}
\end{proposition}

\begin{figure}[H]
	\centering
	\captionsetup{justification=centering}
	\hspace*{-4cm} 
	\begin{tikzpicture}
	[ declare function={binom(\k,\n)=\k!/(\n!*(\k-\n)!);}]
	\begin{axis}[
	axis lines = center,
	axis line style = {->},
	ylabel={$\dfrac{\alpha_{n3,1}}{1-\alpha_{n3,0}}$},
	every axis y label/.style={at={(axis description
			cs:-.5,.5)},anchor=west, rotate=0},
	xlabel={$n$},
	xmin=0,xmax=11.75,
	ymin=0, ymax=1.175,
	every axis x label/.style={at={(current axis.right of 
				origin)},anchor=west},
	samples at={1,...,8}
	]
	\addplot[only marks, black]{binom(10-x,2)*x/(binom(10,3)-binom(10-x,3)};
	\addplot[mark=*, black] coordinates {(9,0) (10,0)};
	\end{axis}
	\end{tikzpicture}
	\caption{Fraction of consumers who observe exactly one price, 
	conditional on observing at least one price, as a function of $n$ for 
	$N=10$ and $k=3$.}
	\label{fig:alpha_nk1}
\end{figure}
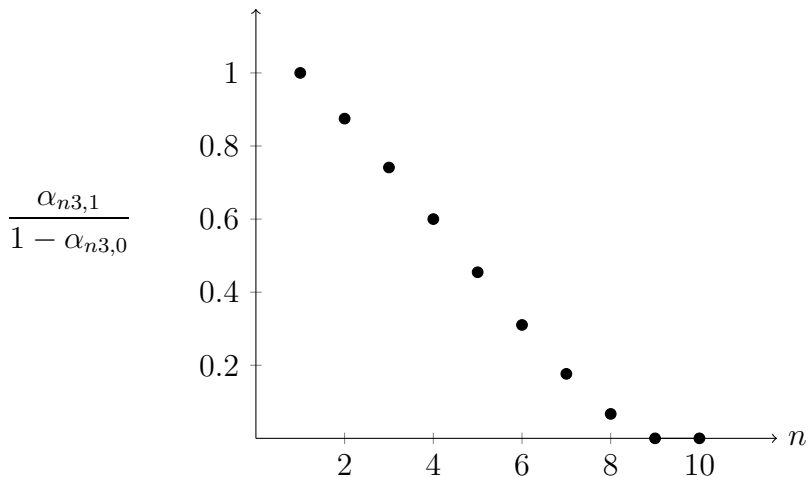

The intuition behind (i) is as follows. The expected price paid 
by a buyer, conditional on observing at least one price, is 
\begin{equation*}
\frac{\alpha_{nk,1}}{1-\alpha_{nk,0}}v.
\end{equation*}
The higher the fraction $\alpha_{nk,1}/(1-\alpha_{nk,0})$, the 
greater the expected price paid. The proof shows that this 
fraction is decreasing in $n$, for $1\leq n\leq N-k+1$. This is 
intuitive as, for a given search strategy, consumers are more 
likely to compare prices when there are more sellers in 
the market.  To illustrate the point, we present Figure 
\ref{fig:alpha_nk1},  which depicts the above fraction (the 
vertical axis) for different values of $n$ (the horizontal 
axis).  However, note that sellers' pricing strategy for each 
realization of $n$ and buyers' search strategy remain the same 
following changes in $\theta$.  Then, more product availability, 
as in (a), translates into a lower share of buyers who drop out 
of the market. This---along with the fact that buyers pay a 
lower expected price conditional on making a purchase---implies 
that buyers' well-being rises. However, this is not true if 
greater product availability is as in (b). In that case, all 
buyers in both markets---one with $i$ number of sellers and the 
other with $j $ number of sellers---make a purchase and pay a 
price equal to the marginal cost of production. Hence, the 
changes in $\theta_i$ and $\theta_j$ do not affect market 
outcomes. In 
Figure \ref{fig:alpha_nk1}, this is illustrated by $j=9$ and 
$i=10$.%

Part (ii) of the proposition is straightforward. An increase in 
search cost causes nothing but a rise in the total resources 
spent on search by consumers. As a result of this, buyers' 
welfare declines.

So far we have considered the impact of marginal changes in 
$\theta$ on market outcomes. Thus, to conclude the section, we 
provide some insights into the impact of substantial (as opposed 
to marginal) changes in product availability on market outcomes. 
From Section \ref{s:BNE}, it follows that there may be two 
stable BNEs for certain parameter regions. One of the BNEs 
occurs in pure strategies of consumers, whereas the other occurs 
in consumers' mixed strategies. Using numerical simulations, we 
report how market outcomes change in those two equilibria.

\begin{figure}[!htb]
	\centering
	\captionsetup{justification=centering}  
	\minipage{0.48\textwidth} 
	\centering
	\includegraphics[width=\linewidth]{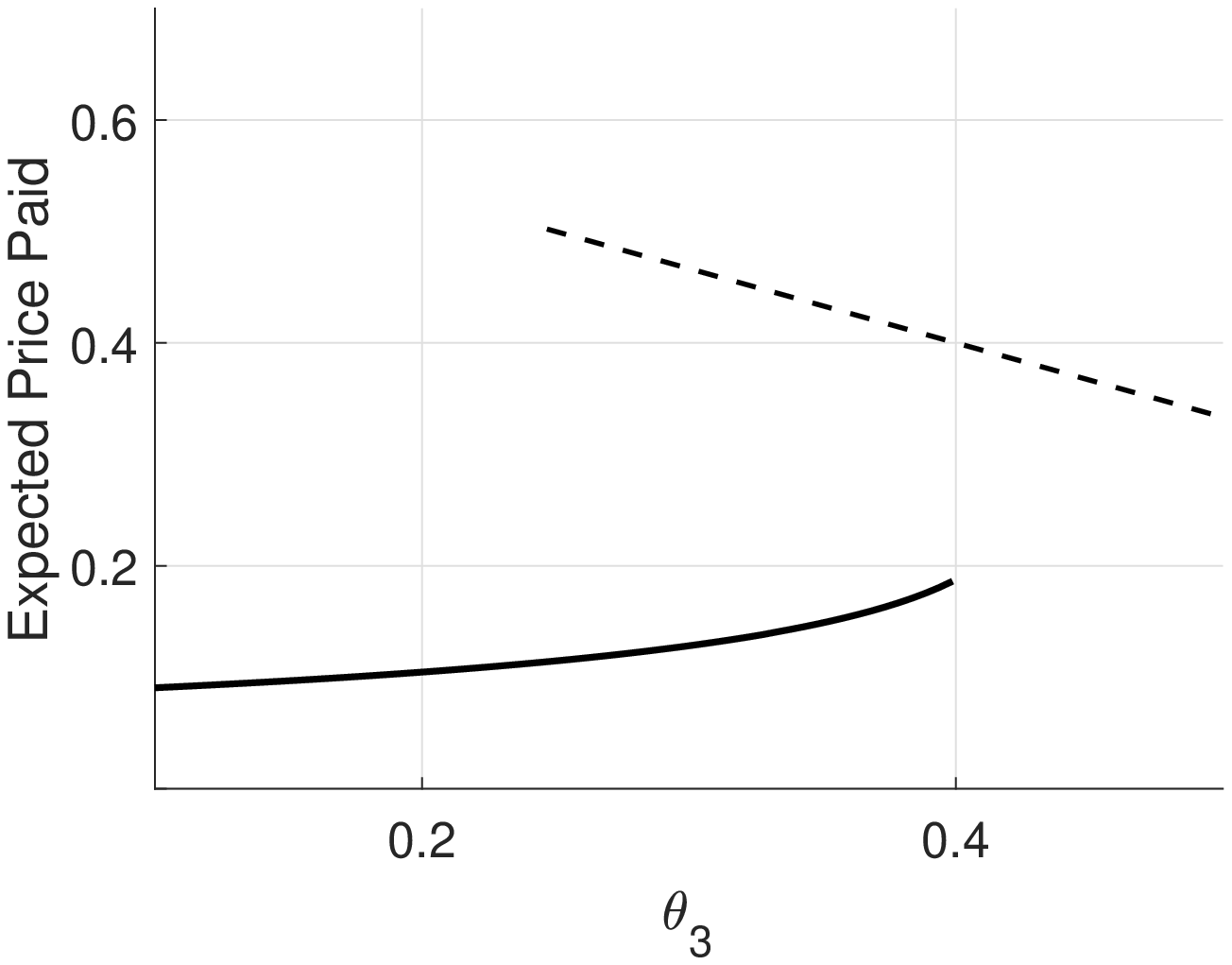}    
	\caption{Impact of greater product availability on the 
	expected price paid.}
	\label{fig:cs_t_Ep}
	\endminipage
	\hfill  
	\minipage{0.48\textwidth}  \centering
	\includegraphics[width=\linewidth]{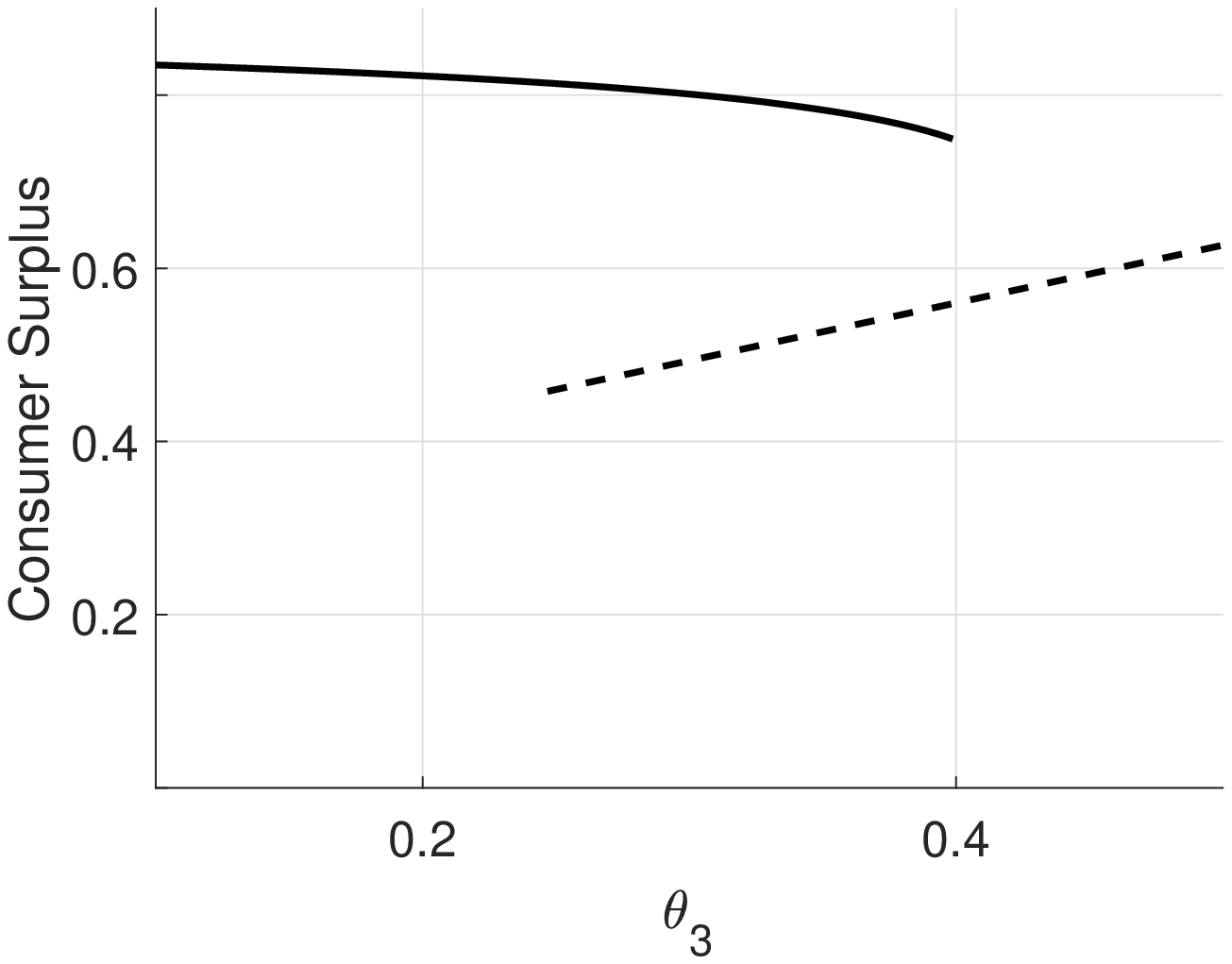} 
	\caption{Impact of greater product availability on the 
	consumer surplus.}
	\label{fig:cs_t_CS}   
	\endminipage
	\hfill
\end{figure}

Figures \ref{fig:cs_t_Ep} and \ref{fig:cs_t_CS} illustrate the 
impacts of greater product availability on the expected price 
paid by buyers and their well-being.  We used the following 
parameter constellations: $N=3$, $\theta_2+\theta_3=1$, $v=1$ 
and $c=0.04$.  The horizontal axes in the figures represent the 
value of $\theta_3$, which we increase from $0.1$ to around 
$0.5$. The vertical axis in Figure \ref{fig:cs_t_Ep} stands for 
the expected price paid by a random buyer, while in Figure 
\ref{fig:cs_t_CS} it stands for buyers' surplus. The solid lines
represent the respective variables in a BNE where buyers play 
mixed strategies. The dashed lines correspond to the respective 
variables in a BNE characterized by pure strategies of buyers.

From both figures we can see that for sufficiently small values 
of $\theta_3 $, there only exists a stable equilibrium in mixed 
strategies of consumers, with some buyers searching two firms and
the remaining ones searching three firms. In contrast, for 
moderately high values of $\theta_3$, only a stable equilibrium 
in pure strategies of buyers exists: all buyers search two 
firms. Finally, for moderate values of $\theta_3$, both stable 
equilibria exist.  

In Figure \ref{fig:cs_t_Ep}, the expected price paid rises with 
$\theta_3$ in the equilibrium where consumers play a mixed 
strategy, whereas the expected price falls with $\theta_3$ in 
the other equilibrium. Notice that for the value of $\theta_3$ 
where both equilibria exist, the expected price is higher in the 
equilibrium characterized by buyers' pure strategies than in the 
other equilibrium. Figure \ref{fig:cs_t_CS} depicts a picture 
similar to Figure \ref{fig:cs_t_Ep}. Importantly, consumers' 
surplus is decreasing with greater product availability in the 
equilibrium where buyers play mixed strategies, while it is 
increasing in the equilibrium with buyers' pure strategies. For 
values of $\theta_3$ where both types of equilibria exist, 
buyers are better off in the equilibrium with mixed strategies. 
This is not surprising, as in the equilibrium with consumers' 
mixed strategies they search more and impose more competitive 
pressure on sellers than in the equilibrium with pure strategies.

\section{Extensions}\label{s:extensions}

In this section, we present extensions of our main model. 
Particularly, we introduce some ``noise'' to the search outcome 
so that, for instance, a consumer who searches once may obtain 
information about offers of more than one firm.  Noisy search 
can represent procurement markets, where a government agency 
announces an auction on a platform and the number of potential 
bidders that will actually see the announcement and participate 
in the auction is uncertain.  In subsection \ref{ss:cost_heter}, 
we allow for search cost heterogeneity by introducing consumers 
with zero search cost.  A common interpretation of such 
consumers is that they use price comparison websites.  Finally, 
in subsection \ref{ss:sequential}, we discuss the role of 
product availability in sequential search markets.

\subsection{Noisy Search}\label{ss:noisy}

For this subsection, we modify our main model by considering 
\textit{noisy search} protocol as in \cite{wilde1977}.  With 
noisy search a buyer faces $N$ different search technologies 
numbered $1,2,...,N$.  Search technology $l \in 
\{1,...,N\}$ entails cost $(l-1)c$ and provides information 
about product availability and prices (if the product is 
available) of $k$ number of firms with probability $\delta^l_k$ 
so that $\sum_{k=l}^N\delta^l_k=1$.  Notice that if 
$\delta^l_l=1$, noisy search collapses to simultaneous search as 
in our main model.  

We make the following assumptions:
\begin{equation}\label{eq:delta}
\begin{aligned}
	&\delta^l_k=0,\ \mbox{for}\  k<l,\\
	&\sum_{k=1}^m\delta^l_k \geq \sum_{k=1}^m\delta^{l+1}_k, \ 
	\mbox{for}\  m \geq l \ \mbox{and} \ 1\leq l \leq 
	N-1,\\
	& \sum_{k=l}^N\left(\delta^{l-1}_k + \delta^{l+1}_k - 2 
	\delta^{l}_k \right) \alpha_{nk}(x) > 0, \ 
	\mbox{for}\  2\leq l \leq N-1, n\geq 2, x\in[0,1].
\end{aligned}
\end{equation}
Assumptions in the first two lines reflect features of 
real-world markets fairly well.  The first line means that a 
buyer searching according to search technology $l$ receives 
information about at least $l$ firms.  The second line implies 
that a higher-numbered search technology is more likely to yield 
information about more firms than a lower-numbered search 
technology.  The assumption in the third line is purely 
technical, and we need it to establish the existence of an 
equilibrium akin to one in Section \ref{ss:kk+1}.  

We do not establish the uniqueness of an equilibrium, since our 
aim is to demonstrate that our results concerning the impact of 
greater product availability on market outcomes are robust to 
our modeling assumptions.  We simply show the existence of an 
equilibrium similar to the unique (stable) one in the main body 
of the paper.  Specifically, we demonstrate that for 
sufficiently small search costs there exists an equilibrium 
where consumers randomize over choosing two adjacent numbers of 
search technologies: $l (=N-\underline{n}+1)$ and 
$l+1(=N-\underline{n}+2)$.  In this equilibrium, sellers price 
the product in a way similar to Proposition 
\ref{prop:eq_k_k+1}.  More specifically, a monopolist seller 
charges price $v$, sellers in a market with $\underline{n}$ 
number of sellers play mixed strategy pricing, and sellers in a 
market with more than $\underline{n}$ number of sellers charge a 
price equal to the production marginal cost.  These observations 
are not surprising since simultaneous search is a special case 
of noisy search.

Since players' equilibrium behavior with noisy search is similar 
to that with simultaneous search, the impact of greater product 
availability on market outcomes is qualitatively the same as in 
the model with simultaneous search.

\begin{proposition}\label{prop:cs_noisy}
	For any $N\geq 3$, $v>0$, $\underline{n}\geq 2$, 
	$\theta_0+\theta_1<1$ and sufficiently small search cost 
	$c>0$, there exists a BNE where consumers randomize over 
	choosing search technologies $N-\underline{n}+1$ and 
	$N-\underline{n}+2$.  
	
	Furthermore, greater product availability 
	associated with an increase in $\theta_i$ and an equal 
	decrease in $\theta_j$ where $i>j$ 
	\begin{itemize}
		\item[(i)] causes weakly more search, reduces the 
		expected price paid and 
		improves consumers' well-being for $j=1$,
		\item[(ii)] does not affect consumers' search 
		behavior, the expected price paid or 
		consumers' well-being for $j\geq \underline{n}+1$,
		\item[(iii)] causes less search, raises the expected 
		price paid and harms consumers' well-being for 
		$j=\underline{n}$.
	\end{itemize}
\end{proposition}

The proof is presented in the appendix.  We omit the detailed 
discussion of the intuition, as it is similar to that of 
Proposition \ref{prop:cs_t_mix}.

\subsection{Search Cost Heterogeneity}\label{ss:cost_heter}

In this subsection, we extend our model to address consumer 
heterogeneity. There are numerous ways to model consumer 
heterogeneity, but here we focus on heterogeneity of search 
costs. There is ample empirical evidence suggesting that 
buyers differ in terms of their search costs in real-world 
markets (e.g., \cite{hongshum2006}, \cite{delossantosetal2012}, 
\cite%
{honkachintagunta2017}). 

One common way of incorporating search cost heterogeneity 
into our model is the introduction of consumers who observe all 
prices in the market for free. With respect to search costs, we 
can think of these buyers as ones who use price comparison 
websites which list all available offers and prices (e.g., 
\cite{varian1980}, \cite{stahl1989}, \cite{janssenmoraga2004}). 
Thus, we assume that $\lambda\in (0,1)$ is the (exogenously 
given) share of buyers who observe all prices. Call them
\textit{costless buyers}, and refer to the rest of the buyers as 
\textit{costly buyers}.  Since our aim is to show that the main 
mechanisms of our model exist in the presence of search 
cost heterogeneity, we will restrict our attention to cases 
where the probability that there are two sellers is strictly 
positive, or $\underline{n}=2$. The rest of the model remains 
unchanged.

As in our main model, there is a unique stable BNE for 
sufficiently small search costs.  In the equilibrium, costly 
buyers randomize over searching $N-1$ and $N$ firms.  Then, it 
is easy to see that a monopolist seller sets price $v$, whereas 
sellers in a market with at least three sellers (namely, $n\geq 
3$) price at the production marginal cost.  Duopoly sellers play 
a mixed strategy. If we let $q\equiv q_{N-1}$ so that 
$q_{N}=1-q$, the expected profit of seller $j$ that sets price 
$p$ is
\begin{equation*}
\Pi_{2j}(p,x_{-j}(p)) = 	\left[(1-\lambda)\left(\frac{q}{N} + 
\left(\frac{q(N-2)}{N} + (1-q)\right)x_2(p)\right) + \lambda x_2(p)\right] p.
\end{equation*}
It is easy to establish that the highest price in the support of 
$x_2(p)$ must be equal to $v$.  Using this fact, we can derive 
the price distribution:
\begin{equation}\label{eq:x2_lambda}
x_2(p) = \mu(q)\left(\frac{v}{p}-1\right), \ \mbox{with support} \ 
\left[\frac{\mu(q)}{1+\mu(q)}v,v\right],
\end{equation}
where
\begin{equation*}
\mu(q) = \frac{q(1-\lambda)}{N-2q(1-\lambda)}
\end{equation*}
which is the ratio of locked-in consumers to that of 
price-comparing buyers (as in traditional models of 
\cite{varian1980} and \cite{stahl1989}).  

Given the above pricing strategies of sellers, a costly buyer is 
indifferent between searching $N-1$ and $N$ firms if
\begin{equation}\label{eq:IC_website}
c =  \frac{2 \theta_2}{N}\left(E[p] - 
E[\min\left\{p_1,p_2\right\}]\right),
\end{equation}
where $E[p]$ is the expected price and 
$E[\min\left\{p_1,p_2\right\}]$ is the expected minimum of the 
two prices in a duopoly market.  

Our main concern is the impact of greater product availability 
on prices and costly consumers' well-being.  The following 
proposition states the main result.

\begin{proposition}\label{prop:cs_website}
	For any $v>0$, $N\geq 3$, $\underline{n}=2$ and $\lambda \in 
	(0,1)$, there exists $\overline{c}>0$ such that for $c\leq 
	\overline{c}$ there exists a unique stable BNE given by 
	$(\left(x_n\right)_{n=1}^N, q)$, where $p=v$ for $n=1$, 
	$p=0$ for $n\geq 3$, $x_2$ is given by 	
	\eqref{eq:x2_lambda}, and $q$ is implicitly determined 
	by \eqref{eq:IC_website}.  
	
	Furthermore, an increase in $\theta_i$ with a corresponding 
	decrease in $\theta_j$
	\begin{itemize}
		\item[(i)] causes weakly more search, reduces the 
		expected price paid and improves costly consumers' 
		well-being for $j=1$, 
		
		\item[(ii)] does not affect consumers' search behavior, 
		the expected price paid or costly consumers' well-being 
		for $j\geq 3$, 
		
		\item[(iii)] causes more search, raises the expected 
		price paid and harms costly consumers' well-being for 
		$j=2$.
	\end{itemize}
\end{proposition}

The intuition is similar to that behind Proposition 
\ref{prop:cs_t_mix} and thus we omit it to avoid repetition. The 
only difference is that in the current proposition we report the 
impact of greater product availability on costly buyers' welfare 
only.  Although analyzing the effect of a change in product 
availability on costless consumers would provide us with 
additional insight, such analysis turns out to be intractable.

\subsection{Sequential Search}\label{ss:sequential}

As a final extension of the model, we consider 
\textit{sequential search} protocol.  Sequential search is a 
type of search such that a consumer who has searched $m$ firms
so far, where $0\leq m\leq N$, decides whether (i) to buy at the 
lowest of the observed prices if there are any, (ii) to search 
for one more firm if there are any non-searched firms, (iii) or
to drop out of the market. We assume that buyers can engage in 
sequential search instead of simultaneous search, and the rest 
of the game and its timing are the same as in the main part of 
our paper.  

\cite{diamond1971} shows that if all consumers face positive 
search costs (without uncertainty about product availability), 
the unique equilibrium is one where consumers do not search 
beyond the first firm and sellers price at $v$.  This result 
holds in our model where consumers face uncertainty about 
product availability.  

To obtain an equilibrium with active search and to provide some 
insights into the effect of greater product availability on that 
equilibrium, we consider a special case of sequential search in 
the working paper version of our study (\cite{atayev2019a}). 
There we incorporate a share of consumers with zero search 
costs.  We also assume that a buyer obtains information about a 
random firm and decides whether to search all of the remaining 
firms.  By doing so, we reduce the number of search paths 
available to buyers.  This search protocol is known as 
\textit{newspaper search} (\cite{varian1980}) and is widely used 
in empirical studies examining gasoline markets (e.g., 
\cite{tappata2009}, \cite{chandratappata2011}). Furthermore, we 
restrict our focus to oligopoly markets with at least two 
and at most three sellers.  This helps us to obtain a mixed 
pricing strategy that does not contain any atoms or flat 
regions.  If there were a possibility of a monopolist seller 
being in the market, oligopoly sellers would want to charge the 
monopoly price with a strictly positive probability to signal an
absence of competition.  Finally, to ease analysis even 
further---and in line with the majority of literature on 
sequential consumer search---we impose \textit{passive} 
out-of-equilibrium beliefs.  This means that, if a consumer 
visits a seller and observes a price that is not a part of an 
equilibrium, she believes that other sellers price the product 
according to an equilibrium strategy.

Under these assumptions, we show that greater product 
availability may harm consumers with a positive search cost.  
Yet the mechanism behind this result is different from that in 
our model of simultaneous search, and is as follows.  With each 
additional seller in the market, an individual seller's chance 
of selling to costless consumers falls more rapidly than their 
share of locked-in (costly) buyers.  Thus, sellers' incentive to 
ripoff locked-in buyers by charging higher prices rather than 
competing fiercely for costless buyers rises with more product 
availability (as in \cite{varian1980} and \cite{stahl1989}).  As 
a result, costly buyers pay higher prices in expectation.

\section{Conclusion}\label{s:conclusion}

We see the current paper as being the first to address 
information asymmetry between buyers and sellers on product 
availability in search markets. The results of the paper suggest 
that technological progress that makes product more available to 
buyers or policy intervention that stimulates entry do not 
necessarily benefit buyers. Thus, accounting for information 
asymmetry on product availability may help to better evaluate 
the impact of such technological progress and policy 
interventions on market outcomes.   

We understand that the model is restricted in the sense that it 
considers a homogeneous goods market. In reality, goods are 
differentiated in many markets and buyers compare different 
deals not only on the basis of prices but also other 
products characteristics, such as match value. Therefore, a 
natural extension of the model is an incorporation of horizontal 
product differentiation as in \cite{perloffsalop1985}, 
\cite{andersonetal1992} and \cite{moragaetal2021}.  In these 
models, uncertain product availability can be interpreted as 
inactive firms ``supplying'' a product with zero match value.  
Then, as products become more available, the dispersion of 
offers across sellers may either increase or decrease depending 
on the distribution of the actual products' match values.  For 
instance, if the distribution of the actual match values is 
skewed towards left (right), more product availability reduces 
the probability that buyers will face inactive firms, or 
products with zero match values.  Therefore, from buyers' 
perspective, the resulting distribution of match values is even 
more (less) skewed towards left and the dispersion of offers 
decreases (increases).    As buyers' search intensity depends on 
the distribution of match values (among other things), the 
impact of greater product availability on buyers' incentive to 
search and, thus, on market outcomes is not straightforward.

\pagebreak 

\appendix

\numberwithin{equation}{section}
\numberwithin{figure}{section}
\begin{singlespace}

\section{Hypergeometric Distribution}

In this section, we discuss application of hypergeometric 
distribution and its properties that will be helpful for our 
proofs.  

The distribution is widely applied in audits, e.g., quality 
control, election audit.  Suppose that an auditor randomly 
selects $k$ items from a population of $N$ items and approves 
a product only if no more than $m$ of the selected items are 
defect, or do not meet quality requirements.  Assume that $n$ 
number of items in the population are defect.  As the 
probability that $m$ of the $k$ selected items follows a 
hypergeometric mass function, one can use the function to test 
what the probability with which the auditor approves the product.

We notice that our probability generating function 
$\alpha_{nk}(x)$ is closely related to the Gauss hypergeometric 
function. If 
\begin{equation*}
	_2F_1(a,b;c;x) \equiv \sum_{m=0}^{n}\frac{(a)_m (b)_m}{(c)_m 
	m!}x^m(p)
\end{equation*}
represents the Gauss hypergeometric function, where $(a)_m = 
a(a+1)...(a+m-1)$, then it follows that 
\begin{equation}\label{eq:gauss}
	\alpha_{nk}(x) = 
	\frac{\binom{N-n}{k}}{\binom{N}{k}}{}_2F_1(-n,-k;N-n-k+1;x).
\end{equation}
To obtain the equation, we first note that for any $n\geq 0$ and 
$0\leq m\leq n$ we have
\begin{equation*}
	\begin{aligned}
\binom{n}{m} = \frac{(n-m+1)_m}{m!},
	\end{aligned}
\end{equation*}
so that
\begin{equation*}
	\begin{aligned}
\alpha_{nk}(x) && =&&& \sum_{m=0}^{n} 
\frac{\binom{N-m}{n-m}\binom{n}{m}}{\binom{N}{k}} x(p)^m  = 
\frac{\binom{N-n}{k}}{\binom{N}{k}}\sum_{m=0}^{n} 
\frac{\binom{N-m}{n-m} \binom{n}{m}}{\binom{N-n}{k}} x(p)^m \\
&&=&&&\frac{\binom{N-n}{k}}{\binom{N}{k}} \sum_{m=0}^{n} 
\frac{(N-n - (k-m) + 1)_{k-m} (n-m+1)_m }{(N-n-k+1)_k} 
\frac{k!}{(k-m)!m!}x(p)^m\\
&&=&&&\frac{\binom{N-n}{k}}{\binom{N}{k}} \sum_{m=0}^{n} 
\frac{(N-n - (k-m) + 1)_{k-m} (n-m+1)_m (k-m+1)_m }{(N-n-k+1)_k 
m!} x(p)^m.
	\end{aligned}
\end{equation*}
Next, as
\begin{equation*}
\frac{(N-n - (k-m) + 1)_{k-m} }{(N-n-k+1)_k} = 
\frac{1}{(N-n-k+1)_m},
\end{equation*}
it follows that
\begin{equation*}
\alpha_{nk}(x) = \frac{\binom{N-n}{k}}{\binom{N}{k}} 
\sum_{m=0}^{n} \frac{ (n-m+1)_m (k-m+1)_m }{(N-n-k+1)_m 
	m!} x(p)^m.
\end{equation*}
Finally, we rewrite
\begin{equation*}
\Resize{}{
	\begin{aligned}
(n-m+1)_m (k-m+1)_m &= n(n-1)...(n-m+1)\times k(k-1)...(k-m+1)\\
&= (-n)(-n+1)...(-n+m-1) \times(-k)(-k+1)...(-k+m-1)\\
&= (-n)_m (-k)_m
	\end{aligned}
}
\end{equation*}
to obtain \eqref{eq:gauss}.

In the following lemma, we note a property of 
hypergeometric 
mass function, which we will use for our proofs.

\begin{lemma}
	\label{lem:stoch_dom} 
	For any $n \geq \underline{n}$ and $k \in \{1,...,N-1\}$, it 
	must be that $\sum_{m=0}^{l}\alpha_{nk,m} \geq 
	\sum_{m=0}^{l}	\alpha_{nk+1,m}$ for all $0\geq l\geq k+1$.
\end{lemma}

\begin{proof}
	We prove the lemma with the help of the following 
	four 
	claims.
	
	\begin{claim}\label{claim:alpha}
		For any $n \in \{0,...,N\}$ and $k \in 
		\{1,...,N\}$, it 
		must 
		be that $\alpha_{nk,m}$ is single-peaked in $m$.
	\end{claim}
	\begin{proof}[Proof of Claim \ref{claim:alpha}]
		It suffices to show that once $\alpha_{nk,m}$ 
		starts 
		decreasing in $m$, it never increases with $m$.  
		Formally, 
		it must be that $\alpha_{nk,m}\geq  
		\alpha_{nk,m+1}$ if
		$\alpha_{nk,m-1}\geq \alpha_{nk,m}$ for $m\geq 
		1$. That 
		$\alpha_{nk,m-1}\geq \alpha_{nk,m}$ for $m\geq 1$ 
		means 
		\begin{equation*}
			\frac{\binom{n}{m-1}\binom{N-n}{k-m+1}}{\binom{N}{k}}
			 \geq 
			\frac{\binom{n}{m}\binom{N-n}{k-m}}{\binom{N}{k}}
			 \ 
			\mbox{for} \ m\geq 1,
		\end{equation*}
		which simplifies to
		\begin{equation}\label{eq:single_1}
			\frac{1}{(n-m+1)(k-m+1)} \geq 
			\frac{1}{m(N-n-k+m)} \ 
			\mbox{for}\ m\geq 1.
		\end{equation}
		Then, it is indeed that $\alpha_{nk,m}\geq  
		\alpha_{nk,m+1}$ 
		if
		\begin{equation*}
			\frac{\binom{n}{m}\binom{N-n}{k-m}}{\binom{N}{k}}
			 \geq 
			\frac{\binom{n}{m+1}\binom{N-n}{k-m-1}}{\binom{N}{k}},
		\end{equation*}
		which simplifies to
		\begin{equation*}
			\frac{1}{(n-m)(k-m)} \geq 
			\frac{1}{(m+1)(N-n-k+m+1)}.
		\end{equation*}
		The last inequality holds, as the denominator on 
		its LHS 
		is smaller than that on the LHS of 
		\eqref{eq:single_1} 
		and the denominator on its RHS is greater than 
		that on 
		the RHS of \eqref{eq:single_1}.  This proves that 
		$\alpha_{nk,m}\geq  \alpha_{nk,m+1}$ if 
		$\alpha_{nk,m-1}\geq \alpha_{nk,m}$  for 
		$m\geq 1$, which implies that $\alpha_{nk,m}$ is 
		single- 
		peaked in $m$.  The proof of the claim is 
		complete.
	\end{proof}
	
	\begin{claim}\label{claim:a_0}
		For any $k \in \{1,...,N-1\}$, it must be $\alpha_{nk,0} 
		> \alpha_{nk+1,0}$ for any $\underline{n}\leq n\leq N-k$ 
		and $\alpha_{nk,1} > \alpha_{nk+1,1}$ for $n = N-k+1$.
	\end{claim}
	\begin{proof}[Proof of Claim \ref{claim:a_0}]
		It is straightforward to calculate:
		\begin{equation*}
			\alpha_{nk,0} - \alpha_{nk+1,0} =
			\frac{\binom{N-n}{k}}{\binom{N}{k}} - 
			\frac{\binom{N-n}{k+1}}{\binom{N}{k+1}} = 
			\frac{(N-n)!(N-k-1)!}{N!(N-n-k-1)!}\left(\frac{N-k}{N-n-k}-1\right)>0.
		\end{equation*}
		For $n = N-k+1$, we have
		\begin{equation*}
			\alpha_{nk,1} - \alpha_{nk+1,1} =
			\frac{\binom{k-1}{k-1}\binom{N-k+1}{1}}{\binom{N}{N-k+1}}
			 - 
			\frac{\binom{k-1}{k}\binom{N-k+1}{1}}{\binom{N}{k+1}}
			 = 
			\frac{\binom{k-1}{k-1}\binom{N-k+1}{1}}{\binom{N}{N-k+1}}>0,
		\end{equation*}
		where the second equality is due to the fact that 
		$\binom{k-1}{k}=0$.
	\end{proof}
	
	\begin{claim}\label{claim:a_unimodal}
		For any $1\leq k \leq N-1$, it must be
		$\argmax\limits_m \alpha_{nk,m} \leq 
		\argmax\limits_m \alpha_{nk+1,m}$.
	\end{claim}
	
	\begin{proof}[Proof of Claim \ref{claim:a_unimodal}]
		Note that $\alpha_{nk,m}$ is strictly positive for 
		$0\leq m \leq \min\left\{n,k\right\}$ and is equal to 
		zero for the other values of $m$.  Also $\alpha_{nk,m}$ 
		is single-peaked, which follows from Claim 
		\ref{claim:alpha}.  Then, $\alpha_{nk,m}$ 
		achieves 
		its maximum at the integer above value $t_k$ 
		which satisfies $\alpha_{nk,t_k} = 
		\alpha_{nk,t_{k+1}}$.  Clearly, the 
		equality implies	
		\begin{equation*}
			\binom{N-n}{k-t_k}\binom{n}{t_k} = 
			\binom{N-n}{k-t_k-1}\binom{n}{t_k+1}.
		\end{equation*}	
		It is easy to check that this can be reduced to
		\begin{equation*}
			(N-n-(k-t_k-1))(t_k+1) = (k-t_k)(n-t_k),
		\end{equation*}
		which yields
		\begin{equation}\label{eq:mk}
			t_k = \frac{(k+1)(n+1)}{N+2}-1.
		\end{equation}
		Notice that it is possible that $t_k<0$, and thus define 
		$M(z)$ to be an operator which spits out $z$ if $z$ is a 
		positive integer, the next high positive integer if $z$ 
		is a positive fraction, and $0$ if $z<0$.  Then, the 
		solution to $\argmax_m \alpha_{nk,m}$ is $t_k$ such that
		\begin{equation*}
			t_k = M(t_k).
		\end{equation*}
		Similarly, the solution to $\argmax_m \alpha_{nk+1,m}$ is
		\begin{equation*}
			t_{k+1} = M(t_{k+1}),
		\end{equation*}
		where, it is easy to check that,
		\begin{equation}
			t_{k+1} = \frac{(k+1)(n+1)}{N+2}.
		\end{equation}
		Then, $\argmax_m \alpha_{nk,m} \leq \argmax_m 
		\alpha_{nk+1,m}$ holds if
		\begin{equation*}
			t_k = \frac{(k+1)(n+1)}{N+2}-1\leq 
			\frac{(k+1)(n+1)}{N+2} =t_{k+1},
		\end{equation*}
		which is certainly true.  This completes the 
		proof of 
		the claim.
	\end{proof}
	
	\begin{claim}\label{claim:d}
		For any $k \in \{1,...,N-1\}$, it must be 
		$\dfrac{\alpha_{nk,m+1}}{\alpha_{nk,m}} < 
		\dfrac{\alpha_{nk+1,m+1}}{\alpha_{nk+1,m}}$ for any 
		$0\leq m \leq k$.
	\end{claim}
	
	\begin{proof}[Proof of Claim \ref{claim:d}]
		It is easy to calculate that
		\begin{equation*}
			\begin{aligned}
				\frac{\alpha_{nk,m+1}}{\alpha_{nk,m}} &&=&&& 
				\frac{\binom{N-n}{k-m-1}\binom{n}{m+1}}{\binom{N-n}{k-m}\binom{n}{m}}
				 = \frac{(k-m)(n-m)}{(N-n-(k-m-1))(m+1)},\\
				\frac{\alpha_{nk+1,m+1}}{\alpha_{nk+1,m}} &&=&&& 
				\frac{\binom{N-n}{k+1-m}\binom{n}{m}}{\binom{N-n}{k-m}\binom{n}{m+1}}
				 = \frac{(k+1-m)(n-m)}{(N-n-(k-m))(m+1)}.
			\end{aligned}
		\end{equation*}
		%		Both $s_{nk,m}$ and $s_{nk+1,m}$ are less than 
		%$1$ for 
		%$m$ close to 
		%		$\min\left\{k,n\right\}$ meaning that the series 
		%are 
		%decreasing in $m$	
		%		for high values of $m$. 
		Then, $\frac{\alpha_{nk,m+1}}{\alpha_{nk,m}} < 
		\frac{\alpha_{nk+1,m+1}}{\alpha_{nk+1,m}}$ is true if
		\begin{equation*}
			\frac{k-m}{N-n-(k-m-1)} < \frac{k+1-m}{N-n-(k-m)},
		\end{equation*}
		which certainly holds because the numerator of the LHS 
		is lower than that of the RHS and the denominator of the 
		LHS is greater than that of the RHS.  The proof 
		of the 
		claim is complete.
	\end{proof}
	
	Claim \ref{claim:alpha} shows that $\alpha_{nk,m}$ is 
	single 
	peaked $m$. Claim \ref{claim:a_0} implies that 
	$\alpha_{nk,m}$ obtains its maximum in $m$ for a 
	weakly 
	lower value of $m$ than $\alpha_{nk+1,m}$.  Claim 
	\ref{claim:a_unimodal} means that when 
	$\alpha_{nk,m}$ is 
	increasing in $m$, it increases more slowly than 
	$\alpha_{nk+1,m}$; $\alpha_{nk,m}$ starts decreasing 
	in $m$ 
	no later than $\alpha_{nk+1,m}$; and when both 
	$\alpha_{nk,m}$ and $\alpha_{nk,m}$ $\alpha_{nk+1,m}$ 
	decrease in $m$, the former decreases faster in $m$ 
	than 
	the latter.  These facts, along with Claim 
	\ref{claim:d} and 
	the fact that $\sum_{m=0}^{k}\alpha_{nk,m}=1$, 
	establish the 
	the proof of the lemma.
\end{proof}

\section{Proofs}

\subsection{Proof of Proposition \ref{prop:pricing}}
	
	For the proof of the proposition, it suffices to show that 
	in any equilibrium 	with active search, the share of buyers 
	who observe two prices is strictly positive for any $n$ 
	where sellers play mixed-strategy pricing.  For the 
	proof that sellers' equilibrium mixed strategy is symmetric 
	and unique if the share of buyers comparing two prices is 
	positive, we refer to \cite{johnenronayne2020}. 
	
	We consider two cases: one where all buyers search the same 
	$k$ number of firms where $2 \leq k\leq N-\underline{n}+1$, 
	and the other where buyers randomize between searching $k$ 
	firms and searching $k+1$ firms where $1\leq k\leq 
	N-\underline{n}+1$. In the former case, the share of 
	consumers that compare two prices is $\binom{N-n}{k-2} 
	\binom{n}{2}/\binom{N}{k}$, which is strictly positive if 
	its numerator is greater than zero.  As $\binom{N-n}{k-2}>0$ 
	for any $\underline{n}\leq n \leq N-k+1$ and $2\leq k \leq 
	N-\underline{n}+1$ and $\binom{n}{2}>0$ for any $n\geq 
	\underline{n}$, the numerator is indeed greater than 
	zero, which means that the share of buyers who observe two 
	prices is strictly positive.  Next, suppose that buyers 
	randomize over searching $k$ and $k+1$ firms where $1\leq k 
	\leq N - \underline{n}+1$.  Then, the share of consumers 
	who compare two prices is a weighted average of 
	$\binom{N-n}{k-2} \binom{n}{2}/\binom{N}{k}$ and 
	$\binom{N-n}{k-1} \binom{n}{2}/\binom{N}{k+1}$.  It is easy 
	to see that the numerator of at least one of these terms is 
	greater than zero for any $\underline{n} \leq n \leq N-k+1$, 
	which shows that the share of buyers comparing two prices is 
	strictly positive.  Thus, in any equilibrium with active 
	search, the share of buyers who observe two prices is 
	strictly positive for any $n$ where sellers play 
	mixed-strategy pricing.  This completes the proof of the 
	proposition.

	\subsection{Proof of Lemma \ref{lem:LHS_IC}}
	
	That $P_k - P_{k+1}$ is positive follows directly from 
	the stochastic dominance in Lemma \ref{lem:stoch_dom}.  
	To show that $P_k - P_{k+1}$ is strictly concave, we 
	note that
	\begin{equation*}
		\Resize{}{
			\dfrac{d (P_k - P_{k+1})}{d q_k} = 
				v \sum_{n=\underline{n}}^{N-k+1} \theta_n 
				\bigintss_{0}^{1} \dfrac{(\alpha'_{nk}(x_n) - 
					\alpha'_{nk+1}(x_n)) \left(\alpha_{nk,1} 
					\alpha'_{nk+1}(x_n) - \alpha_{nk+1,1} 
					\alpha'_{nk}(x_n)\right)}{\left[ 
					\beta'_{nk}(x_n) 
					\right]^2} dx_n,
			}
		\end{equation*}
		and 
		\begin{equation*}
			\Resize{}{
			\dfrac{d^2 (P_k - P_{k+1})}{d q_k^2} = - v 
			\sum\limits_{n=\underline{n}}^{N-k+1} \theta_n 
			\bigintss_{0}^{1} \dfrac{2\beta'_{nk} (x_n) 
			(\alpha'_{nk}(x_n) - \alpha'_{nk+1}(x_n))^2 
			\left(\alpha_{nk,1} \alpha'_{nk+1}(x_n) - 
			\alpha_{nk+1,1} \alpha'_{nk}(x_n)\right) 
			}{\left[\beta'_{nk}(x_n)\right]^4}dx_n,	}
		\end{equation*}
		which is negative if $\Psi_n(x)\equiv \alpha_{nk,1} 
		\alpha'_{nk+1}(x_n) - \alpha_{nk+1,1} \alpha'_{nk}(x_n) 
		\geq 0$ for each $n\geq \underline{n}$ and all $x\in 
		[0,1]$ and $\Psi_n(x)> 0$ for each $n\geq \underline{n}$ 
		and some $x\in [0,1]$.  To simplify the notation, we 
		write $x$ to imply $x_n$ unless stated otherwise. Note 
		that $\Psi_n(0) = \alpha_{nk,1}\alpha_{nk+1,1} - 
		\alpha_{nk+1,1}\alpha_{nk,1} =0$.  Then, $\Psi_n(x)> 0$ 
		for $x\in (0,1]$ if $d \Psi_n(x)/dx >0$ for $x\in 
		(0,1]$.  To show that the derivative is positive, we 
		first note that $\Psi_n(x)$ is $\overline{l}\equiv 
		\min\left\{n-1,k\right\}$ times differentiable in  $x$.  
		Second, we point out that $l$th (where $l\leq 		
		\overline{l}$) derivative of the function is
		\begin{equation}\label{eq:da_dl}
		\begin{aligned}
			\frac{d^l \Psi_n(x)}{d x^l} &&=&&& 
			\alpha_{nk,1}\frac{d^l \alpha'_{nk+1}(x)}{dx^l} 
			- \alpha_{nk+1,1}\frac{d^l\alpha'_{nk}(x)}{dx^l}\\
			&&=&&& 	
			\alpha_{nk,1} \sum_{m=l}^{k+1} \frac{m!}{(m-l)!} 
			\alpha_{nk+1,m}x^{m-l} - \alpha_{nk+1,1}
			\sum_{m=l}^{k}\frac{m!}{(m-l)!}\alpha_{nk,m}x^{m-l}.
			\end{aligned}
		\end{equation}
		Third, we prove that $\frac{d^l \Psi_n(x)}{d x^l}>0$ for 
		each $l$ such that $1\leq l\leq \overline{l}$.  We start 
		by considering $\overline{l}$th  derivative of 
		$\Psi_n(x)$.  For $n-1\geq k$, we have
		\begin{equation*}
			\frac{d^{k} \Psi_n(x)}{dx^{k}} = \alpha_{nk,1}	
			(k+1)!\alpha_{nk+1,k+1} >0.
		\end{equation*}
		This means that $d^{k-1}\Psi_n(x)/dx^{k-1}$ in 
		\eqref{eq:da_dl} is increasing in $x$.  Then, however, 
		it must be that $d^{k-1}\Psi_n(x)/dx^{k-1}>0$ for 		
		$x\in (0,1]$ if $d^{k-1}\Psi_n(x)/dx^{k-1}\geq 0$ for 
		$x=0$. It is easy to see that
		\begin{equation*}
			\left.\frac{d^{k-1} \Psi_n(x)}{d 
			x^{k-1}}\right|_{x=0} = 
			\alpha_{nk,1}k! 
			\alpha_{nk+1,k} - \alpha_{nk+1,1}k!\alpha_{nk,k},
		\end{equation*}
		which is positive if $\alpha_{nk,1}\alpha_{nk+1,k}\geq 
		\alpha_{nk+1,1}\alpha_{nk,k}$, which expands to
		\begin{equation*}
			\frac{\binom{N-n}{k+1-1} \binom{n}{1} 
			\binom{N-n}{k-k} 
				\binom{n}{k}}{\binom{N}{k} \binom{N}{k+1}} \leq	
			\frac{\binom{N-n}{k-1} \binom{n}{1} 
			\binom{N-n}{k+1-k} 
				\binom{n}{k}}{\binom{N}{k} \binom{N}{k+1}}\ 
			\Rightarrow \ \binom{N-n}{k} \leq \binom{N-n}{k-1} 
			(N-n),
		\end{equation*}
		or,
		\begin{equation*}
			\binom{N-n}{k} \leq \binom{N-n}{k} 
			\frac{(N-n)k}{N-n-(k-1)}.
		\end{equation*}
		This is true if $N-n-(k-1)\leq(N-n)k$.  The last 
		inequality can be simplified as $0\leq(N-n+1)(k-1)$ 
		which is clearly true.  This shows that \eqref{eq:da_dl} 
		for $l=k-1$ is positive for $x=0$.  Then, 
		\eqref{eq:da_dl} for $l=k-1$ is strictly positive for 
		any $x\in (0,1]$.

		Now, we repeat similar steps to show that 
		\eqref{eq:da_dl} holds for $l=k-2$.  Namely, the fact 
		that $d^{k-1}\Psi_n(x)/d x^{k-1}>0$ for 
		$x\in (0,1]$ means that $d^{k-2}\Psi_n(x)/d x^{k-2}$ is 
		strictly increasing in $x\in (0,1]$. Then, 
		$d^{k-2}\Psi_n(x)/d x^{k-2}>0$ for all $x\in (0,1]$ if 
		$d^{k-2}\Psi_n(x)/d x^{k-2}|_{x=0}\geq 0.$  Instead of 
		proving the last condition each time, we demonstrate 
		that this holds for each $l$, which is true if 		
		$\alpha_{nk+1,1}l!\alpha_{nk,l} \leq \alpha_{nk,1}l! 
		\alpha_{nk+1,l}$, 
		or
		\begin{equation}\label{eq:alpha_l}
			\alpha_{nk+1,1}\alpha_{nk,l} \leq 
			\alpha_{nk,1}\alpha_{nk+1,l}.
		\end{equation}
		The inequality can be expanded as
		\begin{equation*}
			\frac{\binom{N-n}{k}\binom{n}{1} 
				\binom{N-n}{k-l}\binom{n}{l}}{\binom{N}{k}\binom{N}{k+1}}
				 \leq	
			\frac{\binom{N-n}{k-1}\binom{n}{1} 		
				\binom{N-n}{k+1-l}\binom{n}{l}}{\binom{N}{k}\binom{N}{k+1}},
		\end{equation*}
		which implies 
		\begin{equation*} 
			\binom{N-n}{k} \binom{N-n}{k-l}\leq 
			\binom{N-n}{k-1} 
			\binom{N-n}{k+1-l},
		\end{equation*}
		or,
		\begin{equation*}
			\binom{N-n}{k} 
			\binom{N-n}{k+1-l}\left(\frac{k-l+1}{N-n-(k-l)}\right)
			\leq 
			\binom{N-n}{k} 
			\binom{N-n}{k+1-l}\left(\frac{k}{N-n-(k-1)}\right).
		\end{equation*}
		This reduces to $\frac{k-l+1}{N-n-(k-l)} \leq 
		\frac{k}{N-n-(k-1)}$.  The last inequality clearly holds 
		as the numerator of the LHS is not greater than that of 
		the RHS, and the denominator of the LHS is not smaller 
		than that of the RHS.  This proves that 
		\eqref{eq:alpha_l} holds for any $l$ such that $1\leq l 
		\leq k$, which in its turn proves that \eqref{eq:da_dl} 
		is positive for each $l$, including $l=1$.  Then, it 
		means that $\Psi_n(x)$ is weakly increasing in $x \in 
		[0,1]$ and strictly increasing in $x \in (0,1]$.  Since 
		$\Psi_n(0)=0$, it follows that $\Psi_n(x)>0$ for $x\in 
		(0,1]$.		
				
		Now, it is left to consider the case where $n-1<k$.  
		Like in the previous case, it suffices to show that $d 
		\Psi_n(x)/d x>0$ for $x\in (0,1]$ since $\Psi_n(0)=0$.  
		For that, we apply the same method as for the case of 
		$n-1\geq k$.  First, we note that if each of $l$th 
		($1\leq l\leq n-1$) derivative of $\Psi_n(x)$ with 
		respect to $x$ is positive (and it is strictly positive 
		for $x\in (0,1]$), then $d \Psi_n(x)/d x>0$.  Second, we 
		note that 
		\begin{equation*}
			\frac{d^{n-1}}{d x^{n-1}} \Psi_n(x) =  	
			\frac{\binom{N-n}{k-1} \binom{n}{1} 
			\binom{N-n}{k+1-n} \binom{n}{n}} {\binom{N}{k} 
			\binom{N}{k+1}}n!- \frac{\binom{N-n}{k} \binom{n}{1} 
			\binom{N-n}{k-n} \binom{n}{n}} {\binom{N}{k} 
			\binom{N}{k+1}}n!
		\end{equation*}
		which is strictly positive if 
		\begin{equation*}
			\binom{N-n}{k} 
			\binom{N-n}{k+1-n}\left(\frac{k}{N-n-(k-1)}\right) > 
			\binom{N-n}{k} 
			\binom{N-n}{k+1-n}\left(\frac{k-n+1}{N-k}\right),
		\end{equation*}
		or $\frac{k}{N-n-(k-1)} > \frac{k-n+1}{N-k}$.  Clearly, 
		the inequality holds as the numerator of the LHS is 
		greater than that of the RHS and the denominator of the 
		LHS is smaller than that of the RHS. This demonstrates 
		that $\frac{d^{n-1}}{d x^{n-1}}\Psi_n(x)>0$ for all 
		$x\in [0,1]$.  The next step is to show that 
		$\frac{d^{l}}{d x^{l}}\Psi_n(x)\left.\right|_{x=0}\geq 
		0$ for each $l$ such that $1\leq l \leq n-2$, which 
		proves that $\frac{d^{l}}{d x^{l}}\Psi_n(x)>0$ for $x\in 
		(0,1]$.  The former is certainly true  which follows 
		from the proof of \eqref{eq:alpha_l}.  This proves that 
		$\Psi_n(x)>0$ for $x\in (0,1]$ because 		
		$\Psi_n(0)=0$. =
		
		The proof of the lemma is now complete.

	\subsection{Proofs of Proposition \ref{prop:eq_k_k+1} }
	
	To prove the first part of the proposition, it suffices to 
	show that buyers' participation constraint is satisfied, 
	i.e., buyers prefer searching either $k$ or $k+1$ firms to 
	not searching at all. Fixing pricing strategies of firms in 
	a BNE for $c \in \{\underline{c}_{k,k+1}, 
	\overline{c}_{k,k+1}\}$, we note that an individual buyer's 
	participation constraint is satisfied if $q_k (P_{k} + (k-1) 
	c) + (1-q_k)(P_{k+1} + k c) = P_{k} + (k-1)c <v$, where we 
	used \eqref{eq:IC} to obtain the equality.  This is 
	indeed the case as
	\begin{equation*}
	\begin{aligned}
	P_{k} + (k-1) c &&<&&& P_{k-1} + (k-2) 
	c <...\\
	&&<&&&P_1 = (\theta_0+\theta_1)v + 
	\sum_{n=\underline{n}}^{N-k+1} \theta_n \left(\alpha_{n1,0}v 
	- \alpha_{n1,0} \int_{\underline{p}_n}^{v}px'_n(p)dp\right)\\
	&&<&&&v. 
	\end{aligned} 
	\end{equation*}
	This proves that buyers' participation constraint is indeed
	satisfied, and completes the proof of the first part of the 
	proposition.

	We next prove the second part of the proposition.  For that, 
	it is enough to show that $P_1 - P_2$ and 
	$P_{N-\underline{n}+1} - P_{N-\underline{n}+1}$ converge to 
	zero as $q_1 \uparrow 1$ and $q_{N-\underline{n}+2} \uparrow 
	1$, respectively.  
	
	Since $\alpha'_{n1}(x_n) = \alpha_{n1,1}$, we have
	\begin{equation*}
	\begin{aligned}
	\lim_{q_1 \uparrow 1} (P_1 - P_{2}) &=&& \lim_{q_1 \uparrow 
	1} \sum_{n=2}^N\theta_n v 
	\left((\alpha_{n1,0}-\alpha_{n2,0}) +	
	\int_{0}^{1}\beta_{n1,1} \frac{\left(\alpha'_{n1}(x_n) - 
		\alpha'_{n2}(x_n)\right)}{\beta'_{n1}(x_n) }dx_n\right)\\
	&=&& \sum_{n=2}^N\theta_n v 
	\left((\alpha_{n1,0}-\alpha_{n2,0}) +	
	\int_{0}^{1}\left(\alpha'_{n1}(x_n) - 
	\alpha'_{n2}(x_n)\right)dx_n\right)=0.
	\end{aligned}
	\end{equation*} 
	Here, the first equality in the second line follows from the facts that 
	\begin{equation*}
	\begin{aligned}
	\lim\limits_{q_1\uparrow 1}\beta'_{n1}(x_n) & =&& 
	\lim\limits_{q_1\uparrow 
	1} \left[q_1 \alpha_{n1,1} + (1-q_1)(\alpha_{n2,1} + 
	\alpha_{n2,2}x_n)\right]=\alpha_{n1,1},\\
	\lim\limits_{q_1\uparrow 1}\beta_{n1,1} & =&& \alpha_{n1,1},
	\end{aligned}
	\end{equation*}
	while the second equality in the same line follows from the
	facts that 
	$\alpha_{n1}(1) = \alpha_{n2}(1) = 1$, $\alpha_{n1}(0)=\alpha_{n1,0}$, and 
	$\alpha_{n2}(0)=\alpha_{n2,0}$.  To evaluate 
	$P_{N-\underline{n}+1}-P_{N-\underline{n}+1}$ as 
	$q_{N-\underline{n}+2}\uparrow 1$, we note that, in 
	a BNE where buyers randomize between searching 
	$N-\underline{n}+1(=k)$ and $N-\underline{n}+2(=k+1)$ firms, 
	only sellers in a market with $\underline{n}$ sellers play 
	mixed strategy pricing. Also as $\alpha_{\underline{n}k,0} = 
	\alpha_{\underline{n}k+1,0} = \alpha_{\underline{n}k+1,1} = 
	0$ meaning that $\lim\limits_{q_{k+1}\uparrow 
	1}\beta_{\underline{n}k,1} = 0$ for $k = N-\underline{n}+1$, 
	it follows that 
	\begin{equation*}
	\Resize{}{
		\begin{aligned}
		\lim_{q_{k+1} \uparrow 1} (P_{k} - P_{k+1}) &=\lim_{q_{k+1} \uparrow 1} 
		\theta_{\underline{n}} v \left(\alpha_{\underline{n}k,0} 
		- \alpha_{\underline{n}k+1,0} +	\int_{0}^{1} 
		\frac{\beta_{\underline{n}k,1} 
		\left(\alpha'_{\underline{n}k}(x_{\underline{n}})
		 -\alpha'_{\underline{n}k+1}(x_{\underline{n}})\right)}{ 
		\beta'_{\underline{n}k}(x_2) }dx_{\underline{n}}\right)\\
		&=0.
		\end{aligned}
	}
	\end{equation*}
	Then, it is indeed that $P_1 - P_2$ and 
	$P_{N-\underline{n}+1}-P_{N-\underline{n}+1}$ converge to 
	zero as $q_1 \uparrow 1$ and $q_{N-\underline{n}+2} \uparrow 
	1$, respectively.  This completes the proof of the second 
	part of the proposition.

	\subsection{Proof of Proposition \ref{prop:eq_k}}
	
	We first prove the existence of the cutoff values of the 
	search cost.  In the proof of Proposition 
	\ref{prop:pricing}, we showed that the benefit of searching 
	the $k$th firm is greater than that of searching the $k+1$th 
	firm (given a non-degenerate price distribution(s) for some 
	$n\geq \underline{n}$).  This means that $P_{k-1} + 
	P_{k+1}-2P_k>0$.  Due to strict inequality, it follows that 
	there must a non-empty interval of search costs such that 
	\eqref{eq:IC_2} holds, meaning that $0\leq 
	\underline{c}_k<\overline{c}_k\leq v$.  Furthermore, by 
	construction, it must be that 
	\begin{equation}  \label{eq:cucl}
	\begin{aligned} 
	\lim\limits_{q_k \uparrow 1} (P_k - P_{k+1}) = 
	\underline{c}_k,\\
	\lim\limits_{q_k \uparrow 1} (P_{k-1} - P_k) = 
	\overline{c}_k.
	\end{aligned}
	\end{equation}

	We next show that buyers' participation constraint is 
	satisfied.  This is true if $P_k + (k-1)\overline{c}_k \leq 
	v$.  Given pricing policies of sellers in a BNE for 
	$c = \overline{c}_k$, we have
	\begin{equation*}
	\begin{aligned}
	P_k + (k-1)\overline{c}_k &&<&&& P_{k-1} - 
	(k-2)\overline{c}_k < ...\\
	&&<&&& P_1 = (\theta_0+\theta_1)v + 
	\sum_{n=\underline{n}}^{N-k+1}\theta_n \left(\alpha_{n1,0}v 
	+ 
	\alpha_{n1,1}\int_{\underline{p}_n}^{v}px'_n(p)dp\right)\\
	&&<&&&v.
	\end{aligned}
	\end{equation*}
	Thus, the consumers' participation constraint is indeed 
	satisfied.  This 
	completes the proof of the proposition.

	\subsection{Proof of Corollary \ref{cor:unique_BNE}}
	
	It suffices to prove that (i) $\underline{c}_k>0$ for any $2 
	\leq k \leq N-\underline{n}+1$, (ii) 
	$P_{N-\underline{n}+1}-P_{N-\underline{n}+2}$ is increasing 
	with $q_{N-\underline{n}+1}$ in the neighborhood of $0$ and 
	(iii) $P_{1}-P_{2}$ is decreasing with $q_1$ in the 
	neighborhood of $1$.  
	
	For (i), we note that $\underline{c}_k$ is equal to $P_k - 
	P_{k+1}$ evaluated at $q_k \uparrow 1$ in a BNE where 
	consumers randomize over searching $k$ and $k+1$ firms.  
	Then, it suffices to show that $\lim\limits_{q_k \uparrow 
	1}P_k-P_{k+1}>0$ for any $2\leq k \leq N-\underline{n}+1$.  
	It is easy to see that
	\begin{equation*}
	\lim_{q_k \uparrow 1} (P_k - P_{k+1}) = -\lim_{q_k \uparrow 
	1} \sum_{n=\underline{n}}^{N-k+1}\theta_n 
	\int_{0}^{1}p_n'(x_n)\left(\alpha_{nk}(x_n) - 
	\alpha_{nk+1}(x_n)\right)dx_n.
	\end{equation*} 
	As $p_n'(x_n)<0$, the limiting expression is strictly 
	positive if $\alpha_{nk}(x_n) - \alpha_{nk+1}(x_n)>0$ for 
	some $n$ such that $\underline{n}\leq n \leq N-k+1$.  
	However, for $n = \underline{n}$, the inequality reduces to 
	$\alpha_{\underline{n}k}(x) - \alpha_{\underline{n}k+1}(x) > 
	0$, which certainly holds due to the stochastic dominance in 
	Lemma \ref{lem:stoch_dom}.  Then, the limiting expression is 
	indeed strictly positive. 
	
	For (ii), we recall from Lemma \ref{lem:LHS_IC} that 
	$P_{N-\underline{n}+1} - P_{N-\underline{n}+2}$ is positive 
	and strictly concave in $q_{N-\underline{n}+1}$.  Also from 
	the proof of Proposition \ref{prop:eq_k_k+1}, we 
	know that 
	\begin{equation*}
		\lim\limits_{q_{N-\underline{n}+1} \downarrow 
		0}(P_{N-\underline{n}+1} - P_{N-\underline{n}+2}) =0.
	\end{equation*}
	These two observations mean that $P_{N-\underline{n}+1} - 
	P_{N-\underline{n}+2}$ must be increasing with $q_{N-\underline{n}+1}$ in 
	the neighborhood of $0$.
	
	For (iii), we again recall from Lemma \ref{lem:LHS_IC} that 
	$P_{1} - P_{2}$ is positive and strictly concave in $q_1$.  
	In addition, we know from the proof of Proposition 
	\ref{prop:eq_k_k+1} that
	\begin{equation*}
	\lim\limits_{q_1 \uparrow 1}(P_1 - P_2) =0.
	\end{equation*}
	These two facts imply that $P_1 - P_2$ must be decreasing in 
	$q_1$ in the neighborhood of $1$.
	
	Points (i), (ii), and (iii) establish the proof of the corollary.

	\subsection{Proof of Proposition \ref{prop:cs_t_mix}}

	First, we prove the impact of greater product availability 
	on buyer' search intensity.  For that, we note that $P_k = 
	(\theta_0+\theta_1)v+ \theta_{\underline{n}} 	
	\int_{0}^{1}p_{\underline{n}}(x) \alpha'_{nk}(x)dx$, where 
	$p_{\underline{n}}(x) = q_k 	\alpha_{\underline{n}k,1}v/ 
	\beta'_{\underline{n}k}(x_{\underline{n}})$ and $k= 
	N-{\underline{n}}+1$.  Similarly, $P_{k+1} = 
	(\theta_0+\theta_1)v +  \theta_{\underline{n}} \int_{0}^{1} 
	p_{\underline{n}}(x) \alpha'_{nk+1}(x)dx$.  Therefore, the 
	indifference condition of an individual consumer is given by
	\begin{equation}\label{eq:IC_tau}
		\theta_{\underline{n}} \int_{0}^{1}p_{\underline{n}}(x) 
		\left(\alpha'_{nk}(x)-\alpha'_{nk+1}(x)\right)dx=c.
	\end{equation}
	We know that in equilibrium it must be that 
	\begin{equation}\label{eq:IC_q_mixed}
		\begin{aligned}
	\frac{d \theta_{\underline{n}} 
	\int_{0}^{1}p_{\underline{n}}(x) 
	\left(\alpha'_{nk}(x)-\alpha'_{nk+1}(x)\right)dx}{d 
	\theta_{\underline{n}}} = 
	\int_{0}^{1}p_{\underline{n}}(x)\left(\alpha'_{nk}(x)		
	-\alpha'_{nk+1}(x)\right)dx\\
	+  \theta_{\underline{n}}\frac{\partial 
	\int_{0}^{1}p_{\underline{n}}(x) \left(\alpha'_{nk}(x) - 
	\alpha'_{nk+1}(x)\right)dx}{\partial q_k}\times 		
	\frac{d q_k}{d \theta_{\underline{n}}} = 0.
		\end{aligned}
	\end{equation}
	Note that, as $\int_{0}^{1}p_{\underline{n}}(x) 
	\left(\alpha'_{nk}(x)-\alpha'_{nk+1}(x)\right)dx >0$, it 
	must be that
	\begin{equation*}
		\frac{\partial \int_{0}^{1}p_{\underline{n}}(x) 
		\left(\alpha'_{nk}(x) - 
		\alpha'_{nk+1}(x)\right)dx}{\partial q_k}\times 		
		\frac{d q_k}{d \theta_{\underline{n}}}<0.
	\end{equation*}
	As in the unique stable equilibrium the expected benefit of 
	searching $k+1$th firm is increasing with $q_k$, the partial 
	derivative is positive.  Then, it must be that $d q_k/ d 
	\theta_{\underline{n}}<0$.  From this follows the proof of 
	the effect of greater product availability on buyers' search 
	intensity in all three parts of the proposition. 
	
	Next, we analyze the impact of more product availability on 
	consumer welfare and the expected price paid.  For that, we 
	first observe that for any 	realization of $n>\underline{n}$ 
	sellers price at the production marginal cost in 
	equilibrium.  To see that, replace $k$ by 
	$N-\underline{n}+1$ such that $N-k+2 = 		
	N-(N-\underline{n}+1)+2 = \underline{n}+1$.  From 
	Proposition \ref{prop:eq_k_k+1}, it follows that sellers in 
	a market with at least $N-k+2=\underline{n}+1$ number of 
	sellers price the product at the marginal cost of 
	production.  Parts (i) and (ii) of the proposition directly 
	follows from these facts.
	
	For the proof of part (iii) of the proposition, we note that 
	buyers' total outlay is equal to
	\begin{equation}\label{eq:outlay_mixed}
	q_k P_k + (1-q_k)P_{k+1} + (k+1-q_k)c = (\theta_0+\theta_1)v 
	+ \theta_{\underline{n}} \int_{0}^{1}p_{\underline{n}}(x) 
	\alpha'_{nk+1}(x)dx + (k+1)c,
	\end{equation}
	where we used \eqref{eq:IC_tau} to obtain the equality. The 
	change in the outlay due to \textit{lower} product 
	availability, which is associated with an increase in 
	$\theta_{\underline{n}}$, is given by 
	\begin{equation*}
	\begin{aligned}
	\frac{d\big((\theta_0+\theta_1)v + \theta_{\underline{n}} 
	\int_{0}^{1}p_{\underline{n}}(x) \alpha'_{nk+1}(x)dx + 
	(k+1)c\big)}{d\theta_{\underline{n}}}= \int_{0}^{1}p_{\underline{n}}(x) 
	\alpha'_{nk+1}(x)dx 
	\\
	+ \theta_{\underline{n}} \frac {\partial
	\int_{0}^{1}p_{\underline{n}}(x) \alpha'_{nk+1}(x)dx}{\partial q_k} 
	\times\frac{d q_k}{d \theta_{\underline{n}}}.
	\end{aligned}
	\end{equation*}
	Since in equilibrium \eqref{eq:IC_q_mixed} must hold, the 
	change in buyers' outlay can be rewritten as
	\begin{equation*}
	\int_{0}^{1}p_{\underline{n}}(x) \alpha'_{nk+1}(x)dx
	- \theta_{\underline{n}} \frac{\partial  \int_{0}^{1}p_{\underline{n}}(x) 
	\alpha'_{nk+1}(x)dx}{\partial q_k} \left(\frac{
		\int_{0}^{1}p_{\underline{n}}(x)\left(\alpha'_{nk}(x) - 
		\alpha'_{nk+1}(x)\right)dx}{\theta_{\underline{n}}\frac{\partial
		  \int_{0}^{1}p_{\underline{n}}(x) \left(\alpha'_{nk}(x) 
		- \alpha'_{nk+1}(x)\right)dx}{\partial q_k}}\right),
	\end{equation*}
	which is negative if the following holds
	\begin{equation*}
	\Resize{}{ 
	\dfrac{\partial \int_{0}^{1}p_{\underline{n}}(x) 
	\alpha'_{nk}(x)dx}{\partial q_k}\int_{0}^{1}p_{\underline{n}}(x) 
	\alpha'_{nk+1}(x)dx - \dfrac{\partial \int_{0}^{1}p_{\underline{n}}(x) 
	\alpha'_{nk+1}(x)dx}{\partial q_k}\int_{0}^{1}p_{\underline{n}}(x) 
	\alpha'_{nk}(x)dx<0.
}
	\end{equation*}
	The inequality is true if 
	\begin{equation*}
	\Resize{}{
	\displaystyle{
	\int_{0}^{1}\frac{\alpha_{nk,1}\alpha'_{nk}(x)\alpha'_{nk+1}(x)}
	{\left[\beta'_{nk}(x)\right]^2}dx \int_{0}^{1}\frac{q_k 
	\alpha_{nk,1}\alpha'_{nk+1}(x)}{\beta'_{nk}(x)}dx - 
	\int_{0}^{1}\frac{\alpha_{nk,1}\left[\alpha'_{nk+1}(x)\right]^2}
	{\left[\beta'_{nk}(x)\right]^2}dx \int_{0}^{1}\frac{q_k 
	\alpha_{nk,1}\alpha'_{nk}(x)}{\beta'_{nk}(x)}dx <0,
}	
}
	\end{equation*}
	or
	\begin{equation*}
	\int_{0}^{1}\frac{\alpha'_{nk}(x)\alpha'_{nk+1}(x)} 
	{\left[\beta'_{nk}(x)\right]^2}dx 
	\int_{0}^{1}\frac{\alpha'_{nk+1}(x)}{\beta'_{nk}(x)}dx - 		
	\int_{0}^{1}\frac{\left[\alpha'_{nk+1}(x)\right]^2}			
	{\left[\beta'_{nk}(x)\right]^2}dx 
	\int_{0}^{1}\frac{\alpha'_{nk}(x)}{\beta'_{nk}(x)}dx <0.
	\end{equation*}
	Letting $h \equiv \alpha'_{nk}(x)/\beta'_{nk}(x)$, implying $(1-q_k 
	h)/(1-q_k) = \alpha'_{nk+1}(x)/\beta'_{nk}(x)$, rewrite the inequality as
	\begin{equation*}
	\int_{0}^{1}h(1-q_k h)dx  \int_{0}^{1}(1-q_k h)dx - \int_{0}^{1}(1-q_k 
	h)^2dx \int_{0}^{1}hdx <0.
	\end{equation*}
	This simplifies to
	\begin{equation*}
	-\int_{0}^{1}h^2dx + \left(\int_{0}^{1}hdx\right)^2<0,
	\end{equation*}
	which is clearly true by Cauchy-Schwarz Inequality.  This 
	proves that buyers' outlay decreases with lower product 
	availability; or equivalently, greater product availability 
	harms buyers.  
	
	As a final step, we show that the expected price must 
	increase with greater product availability.  Notice that no 
	buyer drops out of the market for any realization of $n\geq 
	\underline{n}$.  Also following an increase in product 
	availability, buyers economize on their search costs as they 
	search less.  Hence, the only reason why buyers are 
	worse-off due to greater product availability is that the 
	expected price they pay must rise.  This completes the proof.

	\subsection{Proof of Proposition \ref{prop:cs_ck+1}}
	
	Clearly, a change in the search cost does not directly 
	affect sellers.  Hence, any changes in market outcomes 
	caused by the change in the search cost must be through	
	buyers' willingness to search. Recall that, in a stable 
	equilibrium, $P_k-P_{k-1}$ is increasing in $q_k$.  Then, an 
	increase in the search cost, which must be accompanied with 
	an increase in $P_k-P_{k-1}$, must raise $q_k$.
	
	To see that buyers' welfare falls with an increase in $c$, 
	first note that in equilibrium it must be
	\begin{equation}\label{eq:dc}
	\frac{d (P_k - P_{k-1})}{d c} - 1 =0.
	\end{equation}
	Second, the change in the average expected virtual price, 
	denoted by $P$, due to an increase in $c$ can be written as
	\begin{equation*}
	\begin{aligned}
	\frac{d P}{d c} &&=&&& \frac{d \left(P_{k+1} + q_k (P_k - 
		P_{k+1})\right)}{dc} = 
	\left(\frac{\partial P_{k+1}}{\partial q_k} + P_k - 
	P_{k+1}\right)\frac{dq_k}{dc} + q_k\frac{d (P_k - P_{k-1})}{d c}\\
	&&=&&& 	\left(\frac{\partial P_{k+1}}{\partial q_k} + 
	c\right)\frac{dq_k}{dc} + q_k,
	\end{aligned}
	\end{equation*} 
	where the second line is due to \eqref{eq:IC} and 
	\eqref{eq:dc}.  Then, the corresponding change in consumer 
	welfare is
	\begin{equation*}
	\begin{aligned}
	\frac{d(v-P-\left[q_k(k-1) + (1-q_k)k\right]c)}{dc} &&=&&& 
	-\left(\frac{\partial P_{k+1}}{\partial q_k} + 
	c\right)\frac{dq_k}{dc} - q_k - (k-q_k) + \frac{dq_k}{dc}c\\
	&&=&&& - \left(\frac{\partial P_{k+1}}{\partial q_k} 
	\right)\frac{dq_k}{dc} - k.
	\end{aligned}
	\end{equation*}
	As $d q_k/dc>0$, the derivative is negative if $\partial 
	P_{k+1}/\partial q_k \geq 0$, or
	\begin{equation*}
	v \sum_{n=\underline{n}}^{N-k+1} \theta_n \int_{0}^{1} 
	\frac{\alpha'_{nk+1}(x) \left(\alpha_{nk,1} 
	\alpha'_{nk+1}(x) - 	\alpha_{nk+1,1} 
	\alpha'_{nk}(x)\right)}{\left[ \beta'_{nk}(x) \right]^2}dx 	
	\geq 0.
	\end{equation*}
	However, we know from the proof of Proposition 
	\ref{prop:eq_k_k+1} that $\alpha_{nk,1} \alpha'_{nk+1}(x) - 
	\alpha_{nk+1,1} \alpha'_{nk}(x)$ is positive for each $n$ 
	such that $\underline{n}\leq n \leq N-k+1$, which means that 
	the virtual price for $n$ sellers market is indeed 
	increasing in $q_k$.  This shows that $d P_{k+1}$ is 
	increasing in $q_k$, meaning that the derivative of consumer 
	welfare w.r.t. $c$ is decreasing.
	
	The proof is complete.

	\subsection{Proof of Proposition \ref{prop:cs_thk}}
	
	(i) We prove that the average expected  price paid 
	conditional on buying decreases as the product becomes more 
	available. This, along with the fact that with greater 
	product availability the share of consumers who does not 
	make purchase (because they do not find the product) 
	decreases, will mean that buyers' welfare improves when 
	product becomes more available.
	
	Suppose buyers search $k$ firms.  Then, the expected price 
	paid by buyers conditional on their observing at least one 
	price in a market with $n$ sellers is (recall \eqref{eq:xk})
	\begin{equation*}
	\frac{\alpha_{nk,1}}{1-\alpha_{nk,0}}v.
	\end{equation*}
	 To prove (a), we need to show that the fraction 
	$\alpha_{nk,1}/(1-\alpha_{nk,0})$ is decreasing in $n$ such 
	that $1\leq n \leq N-k+1$, or
	\begin{equation*}
	\frac{\alpha_{nk,1}}{1-\alpha_{nk,0}} - 
	\frac{\alpha_{n+1k,1}}{1-\alpha_{n+1k,0}}>0.
	\end{equation*}
	Observe that the inequality certainly holds for 
	$\alpha_{nk,1}\geq \alpha_{n+1k,1}$ because $\alpha_{nk,0} > 
	\alpha_{n+1k,0}$ (which is easy to check).  
	
	Assume that $\alpha_{nk,1}< \alpha_{n+1k,1}$.  Using the 
	definition of $\alpha_{nk,m}$ and simplifying, it is easy to 
	show that $\alpha_{nk,1} < \alpha_{n+1k,1}$ translates into 
	$N-nk-k+1>0$.  Next, simplify the inequality to be proven as
	\begin{equation*}
	\frac{(1-\alpha_{n+1k,0})\alpha_{nk,1} - (1 - \alpha_{nk,0}) 
	\alpha_{n+1k,0}} {(1 - \alpha_{nk,0}) (1 - \alpha_{n+1k,0})} 
	> 0.
	\end{equation*} 
	The inequality holds if the numerator of its left-hand side is positive:
	\begin{equation*}
	(1-\alpha_{n+1k,0})\alpha_{nk,1} - (1-\alpha_{nk,0})\alpha_{n+1k,0}>0.
	\end{equation*}
	Employing the definition of $\alpha_{nk,m}$ expand the LHS of the 
	inequality as follows:
	\begin{equation*}
	\left[1 - \frac{\binom{N-n-1}{k}}{\binom{N}{k}}\right] 
	\frac{\binom{N-n}{k-1}n}{\binom{N}{k}} - 	\left[1 - 
	\frac{\binom{N-n}{k}}{\binom{N}{k}}\right] 
	\frac{\binom{N-n-1}{k-1}(n+1)}{\binom{N}{k}}>0,
	\end{equation*}
	or
	\begin{equation*}
	\left[\binom{N}{k} - \binom{N-n-1}{k}\right] \binom{N-n}{k-1}n - 	
	\left[\binom{N}{k} -\binom{N-n}{k}\right] 
	\binom{N-n-1}{k-1}(n+1)>0,
	\end{equation*}
	Since 
	\begin{equation*}
	\binom{N-n-1}{k} = \binom{N-n}{k}\frac{N-n-k}{N-n},\ \ \ 
	\binom{N-n}{k-1} = \binom{N-n-1}{k-1}\frac{N-n}{N-n-k-1},
	\end{equation*}
	simplify the inequality as
	\begin{equation*}
	-\binom{N}{k}(N-nk-k+1) + 
	\binom{N-n}{k}\left(N-k+1\right)>0.
	\end{equation*}
	Divide both sides of the inequality by $N-nk-k+1(>0)$ and 
	$N-k+1$ and rearrange to obtain
	\begin{equation*} 
	\frac{\binom{N-n}{k}}{N-nk-k+1}>\frac{\binom{N}{k}}{N-k+1},
	\end{equation*}
	or
	\begin{equation}\label{eq:Ep_buy}
	\frac{(N-n)!}{(N-n-k)!(N-nk-k+1)} > \frac{N!}{(N-k)!(N-k+1)}.
	\end{equation}
	
	Observe that for $n=0$, the LHS and the RHS of the 
	inequality are equal to each other.  Then, the inequality 
	holds for all $1\leq n\leq N-k+1$ if the LHS is increasing 
	in $n$.  To show that, we take the derivative of the LHS 
	with respect to $n$ and show that it is positive.  As $n$ is 
	an integer, we apply Gamma function to take the derivative.  
	First, we rewrite the LHS as
	\begin{equation*}
	\frac{\Gamma(N-n+1)}{\Gamma(N-n-k+1)(N-nk-k+1)},
	\end{equation*}
	where $\Gamma$ stands for Gamma function such that 
	$\Gamma(x+1) = x!$.  Noting that, for positive integer $x$,
	\begin{equation*}
	\frac{d \Gamma(x+1)}{dx} = x!\left(-\gamma + 
	\sum_{l=1}^{x}\frac{1}{l}\right),
	\end{equation*}
	where $\gamma = \lim_{x\to \infty}\left(-\ln(x) + 
	\sum_{l=1}^{x}\frac{1}{l}\right)$ is the Euler-Mascheroni 
	constant, the derivative of the LHS of the inequality is
	\begin{equation*}
	\Resize{}{
		\frac{(N-n-k)!(N-n)! \left\{-(N-nk-k+1) \left(-\gamma + 
		\sum\limits_{l=1}^{N-n} \frac{1}{l}\right) + (N-nk-k+1) 
		\left(-\gamma + \sum\limits_{l=1}^{N-n-k} 
		\frac{1}{l}\right) + k\right\}} {\left[\Gamma(N-n-k+1) 
		(N-nk-k+1)\right]^2}.
	}
	\end{equation*}
	The derivative is positive if the term in the curly brackets 
	in the numerator is positive, or $k - (N-nk-k+1) 
	\sum_{l=N-n-k+1}^{N-n} (1/l) > 0$.  This is true as
	\begin{equation*}
	\Resize{}{
		\begin{aligned}
		\frac{k}{N-nk-k+1} &&>&&&\frac{k}{N-n-k+1}= \frac{1}{N-n-k+1} + 
		\frac{1}{N-n-k+1}+...+\frac{1}{N-n-k+1}\\
		&&>&&& \frac{1}{N-n-k+1} + 
		\frac{1}{N-n-k+2}+...+\frac{1}{N-n} = \sum_{l=N-n-k+1}^{N-n}\frac{1}{l}.
		\end{aligned}
	}
	\end{equation*}
	This demonstrates that the derivative of the LHS of 
	\eqref{eq:Ep_buy} is positive.  This, in its turn, implies 
	that \eqref{eq:Ep_buy} is true for $n\geq 1$ and 
	$N-nk-k+1>0$, meaning that the expected price paid 
	conditional on observing at least one price is decreasing 
	with $n$ for $\alpha_{nk,1}<\alpha_{n+1k,1}$.  Then, the 
	average expected price paid conditional on buying is 
	decreasing with greater product availability. 
	
	To show that buyers' welfare increases with greater product 
	availability, it suffices to demonstrate that the expected 
	virtual price falls as the product becomes more available.  
	The latter statement is true if the expected virtual price 
	in a market with $n$ sellers, $P_k^n$, decreases with $n$ 
	for $1\leq n \leq N-k+1$.  Observe that
	\begin{equation*}
	P_k^n = \alpha_{nk,0} v + (1-\alpha_{nk,0}) 
	\frac{\alpha_{nk,1}}{1-\alpha_{nk,0}}v,
	\end{equation*}
	or the expected virtual price in a market with $n$ sellers 
	is a weighted average of the the monopoly price $v$ and the 
	expected price paid by buyers conditional on observing at 
	least one price.  First, it is easy to check that 
	$\alpha_{nk,0}$ is decreasing in $n$.  This means that 
	monopoly price receives less weight while the expected price 
	conditional on purchase receives more weight as $n$ rises.  
	Second, it has been proven above that 
	$\alpha_{nk,1}/(1-\alpha_{nk,0})$ decreases with $n$ such 
	that $1\leq n \leq N-k+1$.  Then, these two effects must 
	clearly cause a decrease in $P_k^n$ as $n$ rises.  This 
	proves that the virtual expected price falls, or that 
	buyers' welfare rises, with greater product availability.  
	
	For (b), to prove that the expected price conditional on 
	observing at least one price does not change with grater 
	product availability for $j\geq N-k+2$, we need to 
	demonstrate that
	\begin{equation*}
	\frac{\alpha_{nk,1}}{1-\alpha_{nk,0}} - 
	\frac{\alpha_{n+1k,1}}{1-\alpha_{n+1k,0}}=0
	\end{equation*}
	for $n\geq N-k+2$.  The equality holds as it is easy to 
	check that $\alpha_{nk,1}=0$ for $n\geq N-k+2$. Using 
	similar steps as in the prove of (a), we can show that 
	$P_k^n$ does not change with $n\geq N-k+2$.  This 
	means that buyers' welfare does not change with greater 
	product availability for $j\geq N-k+2$. 
	
	Proof of (ii) follows directly from discussion after the 
	proposition in the main body of the paper.  
	 
	The proof of the proposition is complete.

	\subsection{Proof of Proposition \ref{prop:cs_noisy}} 
	 
	We first show that for sufficiently small search costs, 
	there exists a stable BNE where some consumers choose 
	search technology $l\geq 2$ for $\underline{n}\geq2$. Then, 
	we prove the impact of greater product availability on 
	market outcomes.
	
	\subsubsection*{Existence of Equilibrium}
	For existence, we first assume that consumers randomize over 
	choosing search technologies $l(=N-\underline{n}+1)$ and 
	$l+1(=N-\underline{n}+1)$, where $1\leq l \leq N-1$ and 
	$\underline{n}\geq2$, and derive pricing policies of 
	sellers.  Later, we show that given the above sellers' 
	pricing decision, an individual consumer is indifferent 
	between choosing search technologies $l$ and $l+1$.  
	
	Thus, suppose that $q$ share of consumers choose search 
	technology $l$ while the remaining consumers choose search 
	technology $l+1$.  Then, a monopolist seller sets price $v$, 
	price in a market with at least $N-l+2$ sellers must be 
	equal to zero, and prices in a market with $\underline{n}$ 
	number of sellers must be dispersed. The equilibrium price 
	distribution in the last market, which we denote by $x$, 
	must satisfy:
	\begin{equation*}
		\sum_{k=l}^{N}\left(q \delta_k^l + 
		(1-q)\delta_k^{l+1}\right) 
		\alpha'_{\underline{n}k}(x(p)) p = 
		\sum_{k=l}^{N}\left(q \delta_k^l + 
		(1-q)\delta_k^{l+1}\right) 
		\alpha'_{\underline{n}k}(x(v))v.
	\end{equation*}
	Here, the LHS is obtained from computing an individual 
	seller's expected profit from setting price $p$ in the 
	support of the equilibrium price distribution $x$.  The 
	RHS of the equation represents the expected profit from 
	selling to only locked-in consumers at a price equal to 
	$v$.  In addition, note that $\alpha'_{\underline{n}k}=0$ 
	for $k>\underline{n}$.  Since the share of consumers who 
	observe two prices is strictly positive, there is a unique 
	solution in $x$ that satisfies the above equation (see, 
	\cite{johnenronayne2020}).
	
	As $x(v)=0$, $\sum_{k=l}^N \delta_k^{l} 
	\alpha'_{\underline{n}k}(0) = 
	\delta_l^l \alpha_{\underline{n}l,1}$ and $\sum_{k=l}^N 
	\delta_k^{l+1} \alpha'_{\underline{n} k}(0)=0$, the inverse 
	function $p(x)$ (in equilibrium) can be written as
	\begin{equation*}
		p(x) = \frac{q \delta_l^l 
		\alpha_{\underline{n}l,1}v}{\sum_{k=l}^{N}\left(q 
		\delta_k^l + (1-q)\delta_k^{l+1}\right) 
		\alpha'_{\underline{n}k}(x)}.
	\end{equation*}
	
	Next, we show that, given the above pricing policies by 
	sellers and a sufficiently small search cost, consumers 
	indeed find it optimal to randomize over choosing search 
	technologies $l=N-\underline{n}+1$ and $l+1= 
	N-\underline{n}+2$.  This is certainly the case if an 
	individual consumer is indifferent between those two 
	options and prefer those options to any other search 
	technology.  To show that, we begin by proving that, for any 
	non-degenerate price distribution for some $n\geq 
	\underline{n}$, consumers cannot be indifferent over 
	choosing $l-1$, $l$ and $l+1$ search technologies where 
	$2\leq l\leq N-1$.  As search technology $N-\underline{n}+1$ 
	yields an expected virtual price as follows
	\begin{equation*}
		P_{l} = (\theta_{0} + \theta_1) v + 
		\sum_{n=\underline{n}}^N\theta_{n} 
		\sum_{k=l}^{N}\delta_{k}^l \int_{0}^{1}p(x_n) 
		\alpha'_{\underline{n} k}(x_n) dx_n
	\end{equation*}
	where we used the fact that $\alpha_{\underline{n} 
	\min\{k,\underline{n}\},0}=0$ for $k \geq 
	N-\underline{n}+1$, a consumer is indifferent of choosing 
	any of the three search technologies if
	\begin{equation*}
		\sum_{n=\underline{n}}^N\theta_{n} \sum_{k=l-1}^{N} 
		\left(\delta_{k}^{l+1} + \delta_{k}^{l-1} - 2 
		\delta_{k}^l \right) \int_{0}^{1} p(x) 
		\alpha'_{\underline{n} k}(x_n) 
		dx_n = 0.
	\end{equation*}
	However, we know that the LHS of the equation is strictly 
	positive by assumption in \eqref{eq:delta}.  This shows that 
	it cannot be that, given price dispersion for some $n\geq 
	\underline{n}$, consumers cannot be indifferent of choosing 
	any of three search technologies $l-1,l$ and $l+1$ for 
	$2\leq l \leq N-1$.  However, this also means that, given 
	non-degenerate price distribution for some $n\geq 
	\underline{n}$, consumers can be indifferent over choosing 
	at most two search technologies so that number of those two 
	search technologies are adjacent.

	We next establish that for sufficiently small search costs 
	and the above pricing policies of sellers, consumers are 
	indifferent of choosing between search technologies 
	$l=N-\underline{n}+1$ and $l+1 =N-\underline{n}+2$. 
	Formally, this translates to
	\begin{equation}\label{eq:IC_noisy}
		\theta_{\underline{n}} \sum_{k=l}^{N}\left(\delta_{k}^l -
		\delta_{k}^{l+1}\right) \int_{0}^{1} p(x) 
		\alpha'_{\underline{n}k}(x) dx=c,
	\end{equation}
	where we used the fact that there is only price dispersion 
	in a market with $\underline{n}$ number of sellers and
	consumers in that market observe at least one price, and we 
	wrote simple $x$ to mean $x_{\underline{n}}$.

	We note the following two facts about the LHS of the 
	equation.  First, the LHS is positive for $0<q<1$.  To see 
	that, rewrite the LHS as 
	\begin{equation*}
	\begin{aligned}
	\sum_{k=l}^{N}\left(\delta_{k}^l -
	\delta_{k}^{l+1}\right) \int_{0}^{1} p(x) 
	\alpha'_{\underline{n}k}(x) dx = 
	\sum_{k=l}^{N}\left(\delta_{k}^l -
	\delta_{k}^{l+1}\right)\left(\underline{p}_{\underline{n}} - 
	\int_{0}^{1} p'(x) \alpha_{\underline{n} k}(x) dx\right)\\
	= \underline{p}_{\underline{n}} \sum_{k=l}^{N} 
	\left(\delta_{k}^l -\delta_{k}^{l+1}\right) - \int_{0}^{1} 
	p'(x) \sum_{k=l}^{N}\left(\delta_{k}^l - 
	\delta_{k}^{l+1}\right) \alpha_{\underline{n} k}(x) dx.
	\end{aligned}
	\end{equation*}
	Clearly, the first term is positive by the assumption in 
	\eqref{eq:delta}.  The second term is positive if 
	$\sum_{k=l}^{N}\left(\delta_{k}^l -
	\delta_{k}^{l+1}\right) \alpha_{\underline{n} 
	k}(x)\geq 0$. To prove that this 
	inequality holds, we use the following algorithm.  We start 
	at $k=N$.  Three cases are possible: (i) both $\delta_N^l$ 
	and $\delta_N^{l+1}$ are strictly positive, (ii) only 
	$\delta_N^{l+1}$ is strictly positive, or (iii) both  
	$\delta_N^l$ and $\delta_N^{l+1}$ are equal to zero.  In 
	case (i), it must be $\delta_N^{l} < \delta_N^{l+1}$ by 
	assumption.  In this case, we ``take away'' $\tau_{N-1} \in 
	(0,1]$ share of $\delta_{N-1}^{l}$ so that 
	$\tau_{N-1}\delta_{N-1}^l + \delta_{N}^{l} - 
	\delta_{N}^{l+1}=0$, for 
	which we have $\tau_{N-1} \delta_{N-1}^{l} 
	\alpha_{\underline{n} k}(x) + 
	(\delta_{N}^{l} - \delta_{N}^{l+1})
	\alpha_{\underline{n}k}(x) > 0$ due to 
	stochastic dominance in Lemma \ref{lem:stoch_dom}.  If  
	$\delta_{N-1}^{l} \delta_{N}^{l} - \delta_{N}^{l+1} <0$, we 
	in addition take 	$\tau_{N-2}\delta_{N-2}^{l}$ and add it 
	to the LHS of the inequality so that it is zero; and if it 
	is still negative, we repeat the procedure.  As a second 
	step, we undertake similar actions for  $k=N-1$.  We 
	continue such steps until we get to step $k=l+1$.  At this 
	point, there must be left either
	$\delta_{l}^l \alpha_{\underline{n}l}(x) - 
	\left((1-\tau_{l+1})\delta_{l+1}^l +
	\delta_{l+1}^{l+1} \right)\alpha_{\underline{n}l+1}(x)$ with 
	$
	\delta_{l}^l + (1-\tau_{l+1})\delta_{l+1}^l- 
	\delta_{l+1}^{l+1}=0$ or $ (1-\tau_{l}) \delta_{l}^l 
	\alpha_{\underline{n}l}(x) - \delta_{l+1}^{l+1} 
	\alpha_{\underline{n}l+1}(x)$ with $(1-\tau_{l})\delta_{l}^l 
	- \delta_{l+1}^{l+1}=0$, both of which are strictly 
	positive.  In cases (ii) and (iii), we follow the same 
	algorithm as we have laid out for case (i) with an exception 
	that in case (iii) we simply skip step $k=N$.  This proves 
	that $\sum_{k=l}^{N}\left(\delta_{k}^l - 
	\delta_{k}^{l+1}\right) \alpha_{\underline{n} k}(x)\geq 0$, 
	or the LHS of \eqref{eq:IC_noisy} is positive.  

	The second fact is that, as $q \downarrow 0$, the LHS of 
	\eqref{eq:IC_noisy} converges to zero:
	\begin{equation*}
	\begin{aligned}
		&&&\lim\limits_{q \downarrow 0} \theta_{\underline{n}} 
		\sum_{k=l}^{N}\left(\delta_{k}^l -
		\delta_{k}^{l+1}\right) \int_{0}^{1} p(x) 
		\alpha'_{\underline{n}k}(x) dx\\
		=&&& 
		\theta_{\underline{n}} 
		\sum_{k=l}^{N}\left(\delta_{k}^l -
		\delta_{k}^{l+1}\right) \int_{0}^{1} \lim_{q\downarrow 
		0} \left(\frac{q \delta_l^l 
		\alpha_{\underline{n}l,1}v}{\sum_{k=l}^{N}\left(q 
		\delta_k^l + (1-q)\delta_k^{l+1}\right) 
		\alpha'_{\underline{n}k}(x)}\right)
		 \alpha_{\underline{n}k} (x) dx\\
		=&&&0,
	\end{aligned}	
	\end{equation*}
	where we used inverse functions $p(x)$ to obtain the first 
	equality.
	
	These two facts---that the LHS of \eqref{eq:IC_noisy} is 
	positive and converges to zero as $q \downarrow 0$---mean 
	that for sufficiently small search costs there exists 
	$0<q<1$ that solves \eqref{eq:IC_noisy}.  Moreover, this 
	equilibrium is locally stable as the LHS is strictly 
	increasing in $q$ for sufficiently small search costs.  This 
	completes the proof of the existence of a BNE.

	\subsubsection*{Impact of Greater Product Availability}
	
	Now, we turn on proving the impact of greater product 
	availability on buyers' search behavior, average expected 
	price paid by them and their surplus. 
	
	Parts (i) and (ii) of Proposition \ref{prop:cs_noisy} are 
	straightforward,  which is why we only prove part (iii) of 
	the proposition.  In equilibrium we have 
	\begin{equation}\label{eq:IC_q_noisy}
	\frac{d q}{d \theta_{\underline{n}}} = - 
	\frac{\frac{\partial (P_{N-\underline{n}+1} - 	
	P_{N-\underline{n}+2})}{\partial \theta_{\underline{n}}}}{ 
	\frac{\partial (P_{N-\underline{n}+1} - 
	P_{N-\underline{n}+2})}{\partial q}},
	\end{equation}
	which is negative as the numerator of the RHS is clearly 
	positive and the denominator of the RHS is positive
	because of the above facts that the LHS of 	
	\eqref{eq:IC_noisy} (which is nothing but 
	$P_{N-\underline{n}+1} - P_{N-\underline{n}+2}$) is positive 
	for $0<q<1$ and converges to zero as $q \downarrow 0$.  
	Next, consumers' expected outlay is equal to
	\begin{equation*}
		q P_{N-\underline{n}+1} + (1-q)P_{N-\underline{n}+2} + 
		(N-\underline{n}+1-q) c = P_{N-\underline{n}+2} + 
		(N-\underline{n}+1)c,
	\end{equation*}
	where we used the fact that in equilibrium we have 
	$P_{N-\underline{n}+1} - P_{N-\underline{n}+2}=c.$  We need 
	to prove that this outlay is decreasing with 	
	$\theta_{\underline{n}}$ (and the associated equal decrease 
	in $\theta_j$ where $j>\underline{n}$).  This is true if the 
	derivative of the outlay w.r.t. $\theta_n$ is negative.  
	This derivative is equal to
	\begin{equation*}
		\frac{\partial  P_{N-\underline{n}+2}}{\partial 
		\theta_{\underline{n}}} + \frac{\partial  
		P_{N-\underline{n}+2}}{\partial q} \frac{d
		q}{d \theta_{\underline{n}}}.
	\end{equation*}
	Using \eqref{eq:IC_q_noisy}, we can rewrite the derivative 
	of the outlay w.r.t. $\theta_n$ as
	\begin{equation}\label{eq:cs_noisy}
		\frac{\dfrac{\partial P_{N-\underline{n}+1}}{\partial q} 
		\dfrac{\partial P_{N - \underline{n}+2}}{\partial 
		\theta_{\underline{n}}} - \dfrac{\partial 
		P_{N-\underline{n}+1}}{\partial \theta_{\underline{n}}} 
		\dfrac{\partial P_{N - \underline{n}+2}}{\partial 
		q}}{\dfrac{\partial (P_{N-\underline{n}+1} - 
		P_{N-\underline{n}+2})}{\partial q}}.
	\end{equation}
	As the denominator of the expression is positive, 
	the expression in \eqref{eq:cs_noisy} is negative if its 
	numerator is negative.  The numerator is equal to
	\begin{equation*}
		\Resize{}{
	\begin{aligned}
		\theta (\delta_l^l \alpha_{\underline{n}l,1}v)^2 
		\int_{0}^{1}\frac{\sum_{k=l}^{N}\delta_k^l 
		\alpha_{\underline{n} k}'(x) 
		\sum_{k=l}^{N}\delta_k^{l+1} 
		\alpha_{\underline{n} k}'(x)
		}{\left(\sum_{k=l}^{N}\left(q 
		\delta_k^l + (1-q)\delta_k^{l+1}\right) 
		\alpha'_{\underline{n}k}(x)\right)^2}dx 
		\int_{0}^{1}\frac{\sum_{k=l}^{N}\delta_k^{l+1} 		
		\alpha_{\underline{n} 
		\min\{k,\underline{n}\}}'(x)}{\sum_{k=l}^{N}\left(q 
		\delta_k^l + (1-q)\delta_k^{l+1}\right) 
		\alpha'_{\underline{n}k}(x)}dx\\
	 - \theta (\delta_l^l \alpha_{\underline{n}l,1}v)^2
	 	\int_{0}^{1}\frac{\sum_{k=l}^{N}\delta_k^{l} 
	 	\alpha_{\underline{n}k}'(x)}{\sum_{k=l}^{N}\left(q		
	 	\delta_k^l + (1-q)\delta_k^{l+1}\right) 
	 	\alpha'_{\underline{n}k}(x)}dx	 
	 	\int_{0}^{1}\frac{\left(\sum_{k=l}^{N}\delta_k^{l+1} 
	 	\alpha_{\underline{n} k}'(x) 
	 	\right)^2}{\left(\sum_{k=l}^{N}\left(q \delta_k^l + 
	 	(1-q)\delta_k^{l+1}\right) \alpha'_{\underline{n} 
	 	k}(x)\right)^2}dx.
	\end{aligned}
}
	\end{equation*}
	To see that this is negative, let $h 
	\equiv \frac{\sum_{k=l}^{N}\delta_k^{l} 
	\alpha_{\underline{n}k}'(x)}{\sum_{k=l}^{N}\left(q 
	\delta_k^l + (1-q)\delta_k^{l+1}\right) 
	\alpha'_{\underline{n}k}(x)}$
	 so that $(1-qh)/(1-q) = \frac{\sum_{k=l}^{N}\delta_k^{l+1}  
	 \alpha_{\underline{n} k}'(x)}{ 
	 \sum_{k=l}^{N}\left(q \delta_k^l + 	 	
	 (1-q)\delta_k^{l+1}\right) 	
	 \alpha'_{\underline{n}k}(x)}$ as we 
	 did in the proof of Proposition \ref{prop:cs_t_mix}, and 
	 simplify to obtain an expression which is the same as that 
	 on the LHS of the last inequality in the proof of 
	 Proposition \ref{prop:cs_t_mix}, which is clearly 
	 negative.  This means that the numerator of 
	 \eqref{eq:cs_noisy} is  negative, which in turn proves that 
	 the derivative of consumers' outlay w.r.t. 
	 $\theta_{\underline{n}}$ is negative.  As consumers search 
	 more and are better-off due to a decrease in product 
	 availability, associated with an increase in 
	 $\theta_{\underline{n}}$, the expected price paid must be 
	 lower.  The proof of the proposition is now complete.

	\subsection{Proof of Proposition \ref{prop:cs_website}}
	
	We first prove the existence, and later prove the 
	comparative static result.  
	
	\subsubsection*{Existence}
	
	We will use the following facts for the proof.  
	First,  	
	$E[p]=v+\int_{\underline{p}}^{v}(1-x_2(p))dp$. Second, 
	$E[\min\left\{ 
	p_{1},p_{2}\right\}] = 
	v-2\int_{\underline{p}}^{v}(1-x_2(p))dp + 
	\int_{\underline{p}}^{v}(1-x_2(p))^2 dp$, $k\neq l$ while 
	\begin{equation*}
		\begin{aligned}
			E[p]-E[\min \left\{p_{1},p_{2}\right\}]&& =&&&  
			\int_{\underline{p}}^{v}(1-x_2(p))dp - 
			\int_{\underline{p}}^{v}(1-x_2(p))^2dp\\
			&&=&&& v\mu(q)\left(\left(1+2\mu(q)\right)\ln\left(1 
			+ 
			\frac{1}{\mu(q)}\right)-2\right).
		\end{aligned}
	\end{equation*}
	Then, we can rewrite \eqref{eq:IC_website} as 
	\begin{equation}\label{eq:IC_web_high}
		\frac{c}{v} = \theta_2\frac{2}{N}\mu(q)
		\left(\left(1+2\mu(q)\right)\ln\left(1 + 
		\frac{1}{\mu(q)}\right) - 
		2\right).
	\end{equation}
	The RHS of the equation is positive only if 
	$\left(1+2\mu(q)\right)\ln\left(1 + \frac{1}{\mu(q)}\right)  
	> 2$, or $\ln\left(1 + \frac{1}{\mu(q)}\right)  > 
	\frac{2}{1+2\mu(q)}$. Note that when $\mu(q) \downarrow 0$ 
	the LHS of the inequality goes to infinity while its RHS 
	converges to $2$, and when $\mu(q) \to \infty$ both the LHS 
	and the RHS converge to $0$.  Then, it suffices to show that 
	the derivative of LHS is more negative than that of the RHS 
	for the inequality to hold.  Indeed, the derivative of the 
	LHS is $-\frac{1}{\mu(q)(1+\mu(q))} = - \frac{1+4\mu(q) + 	
	4\mu(q)^2}{\mu(q)(1+\mu(q))(1+2\mu(q))^2}$ while that of the 
	RHS is $- \frac{4}{(1+2\mu(q))^2} = - \frac{4\mu(q) + 		
	4\mu(q)^2}{\mu(q)(1+\mu(q))(1+2\mu(q))^2}$, and the former 
	is more negative than the latter.  Summing up, this proves 
	that the RHS of \eqref{eq:IC_web_high} is positive.
	
	Now, we show that the RHS of \eqref{eq:IC_web_high} is 
	inverse U-shaped in $q$, which is true only if it is inverse 
	U-shaped which respect $\mu(q)$ as $\mu(q)$ is increasing in 
	$q$ with $\mu(0)=0$ and $\mu(1) = 
	(1-\lambda)/(N-2+2\lambda)$.  The derivative of the RHS 	
	with respect to $\mu(q)$ is
	\begin{equation*}
		\begin{aligned}
			&&&\frac{2\theta_2}{N} 
			\left(\left(1+2\mu(q)\right)\ln\left(1 + 
			\frac{1}{\mu(q)}\right) - 2 + \mu(q)\left[2 
			\ln\left(1 + 
			\frac{1}{\mu(q)}\right) - \frac{1+2\mu(q)}{\mu(q) + 
				\mu(q)^2}\right]\right)\\
			=&&& \frac{2\theta_2}{N} \left(\frac{(1+5\mu(q)+ 4 
				\mu(q)^2)\ln\left(1 + 
				\frac{1}{\mu(q)}\right) - 3 - 4\mu(q)}{1 + 
				\mu(q)}\right),
		\end{aligned}
	\end{equation*}
	which is equal to zero only if its numerator is zero, or 
	$M(\mu(q)) \equiv \ln\left(1 + \frac{1}{\mu(q)}\right) - 
	\frac{3 + 4\mu(q)}{(1+\mu(q))(1 + 4\mu(q))}=0$.   Thus, if 
	$M(\mu(q))=0$ for only a single $\mu(q)$, then the RHS of 
	\eqref{eq:IC_web_high} has one stationary point in $q \in 
	(0,1)$.  The following facts, along with the fact that 
	$M(\mu(q))$ is continuous in $\mu(q)>0$, prove that 
	$M(\mu(q))=0$ has a single solution in $\mu(q)\in(0,1)$:
	\begin{equation*}
		\begin{aligned}
			&M(0) = + \infty,\  \mbox{and}\  M(\infty) = 0,\\
			&\frac{\partial M(\mu(q))}{\partial \mu(q)} = 
			\frac{2 \mu(q)-1}{\mu(q)(1+\mu(q))^2(1+4\mu(q))^2} 
			\begin{cases*}
				< 0 \ \mbox{for} \ \mu(q) <\frac{1}{2},\\
				= 0  \ \mbox{for} \ \mu(q)=\frac{1}{2},\\
				> 0 \ \mbox{for} \ \mu(q) >\frac{1}{2}.
			\end{cases*}
		\end{aligned}
	\end{equation*}
	Thus, the RHS of \eqref{eq:IC_web_high} has a unique 
	stationary point in $q\in(0,1)$.  To see that the RHS is 
	maximized at that stationary point, note that the derivative 
	of the RHS is positive for some values of $q$ close to $0$ 
	and negative for some values of $q$ close to $1$. This 
	completes the proof that the RHS is inverse U-shaped and has 
	a unique maximum in $q \in (0,1)$.
	
	Next, we demonstrate that the RHS of \eqref{eq:IC_web_high} 
	is less than $1$.  For that, we rewrite the RHS as
	\begin{equation*}
		\begin{aligned}
			\frac{c}{v} = 
			&&&\frac{2}{N}\theta_2\left(\mu(q)\ln\left(1 + 
			\frac{1}{\mu(q)}\right) + 2 
			\mu(q)^2 \ln\left(1 + \frac{1}{\mu(q)}\right) - 
			2\mu(q)\right)\\
			=&&& \frac{2}{N}\theta_2\left\{\mu(q)\ln\left(1 + 
			\frac{1}{\mu(q)}\right) - 
			2 \mu(q)\left[1-\mu(q) \ln\left(1 + 
			\frac{1}{\mu(q)}\right)\right]\right\}.
		\end{aligned}
	\end{equation*}
	Next, we note that the first term in the large brackets, 
	which is positive, is increasing in $\mu(q)$ and converges 
	to $1$ as $\mu(q) \to \infty$.  Since the terms in the 
	square brackets is positive and that the RHS is positive, 
	the RHS must be less than $1$.
	
	Finally,  we show that the RHS of \eqref{eq:IC_web_high} 
	converges to zero as $q\to 0$, or $\mu(q) \to 0$.  Note that 
	since 
	\begin{equation*}
		\begin{aligned}
			\lim\limits_{\mu(q)\to 0}\mu(q) \ln\left(1 + 
			\frac{1}{\mu(q)}\right) &&=&&& 
			\lim\limits_{z \to \infty} \frac{\ln(1+z)}{z} \ \ \  
			\myeq \ 
			\ \ 
			\lim\limits_{z \to \infty} \frac{1}{1+z} = 0,
		\end{aligned}
	\end{equation*}
	we have
	\begin{equation*}
		\lim\limits_{\mu(q) \to 0} \left(\mu(q)\ln\left(1 + 
		\frac{1}{\mu(q)}\right) 
		- 2 \lim\limits_{\mu(q) \to 0} \mu(q)\left[1- 
		\lim\limits_{\mu(q) \to 
			0}\mu(q) \ln\left(1 + 
			\frac{1}{\mu(q)}\right)\right]\right) 
		= 0.
	\end{equation*}
	Therefore, the RHS of \eqref{eq:IC_web_high} converges to 
	zero as $q \to 0$.

	The facts that, for $0<q<1$ the RHS of 
	\eqref{eq:IC_web_high} is positive for, inverse U-shaped in 
	$q$, and converges to zero as $q \to 0$, there 
	must be a stable BNE for sufficiently small search cost.  
	The fact that the RHS \eqref{eq:IC_web_high} is less than 
	one proves that $\overline{c}<v$.
	
	\subsubsection*{Comparative Staics}
	
	Parts (i) and (ii) are straightforward which is why they are 
	omitted.
	
	We prove part (iii). For that, we note that in 
	equilibrium it must be that 
	\begin{equation}\label{eq:dtau_web}
		\frac{d}{d \theta_2}\left[\frac{2}{N}\theta_2\left(E[p]- 
		E[\min\{p_1,p_2\}]\right)\right]=0,
	\end{equation}
	or
	\begin{equation*}
		E[p]- E[\min\{p_1,p_2\}] 
		+\theta_2\frac{\partial{\left(E[p] 
				- 	E[\min\{p_1,p_2\}]\right)}}{\partial \mu(q)} 
				\times 
		\frac{d\mu(q)}{d \theta_2} =0,
	\end{equation*}
	so that
	\begin{equation}\label{eq:dmu_dt}
		\frac{d \mu(q)}{d \theta_2} = - \frac{E[p]- 
			E[\min\{p_1,p_2\}]}{\theta_2\frac{\partial{\left(E[p]
					- E[\min\{p_1,p_2\}]\right)}}{\partial 
					\mu(q)}}.
	\end{equation}
	This derivative is negative as the denominator on the RHS of 
	the equation is positive because we know that in a stable 
	equilibrium $E[p] - E[\min\{p_1,p_2\}]$ is increasing in 
	$\mu(q)$.  As $\mu(q)$ is increasing in $q$, it follows that 
	$q$ is decreasing with  $\theta_2$.
	
	Next, costly buyers are better-off with an increase in 
	$\theta_2$, associated with a decrease in $\theta_j$ for 
	$j\geq 3$, if their outlay is decreasing with $\theta_2$, or
	\begin{equation*}
		\Resize{}{
			\dfrac{d}{d\theta_2}\left[\theta_2\left(q\dfrac{2}{N}E[p]
			 + 
			q \dfrac{(N-2)}{N}E[\min\{p_1,p_2\}]) + (1-q)E 
			[\min\{p_1,p_2\}]\right) + 
			(N-1-q)c \right]<0.
		}
	\end{equation*}
	Using \eqref{eq:IC_website} to substitute the value of $c$, 
	we obtain
	\begin{equation}\label{eq:Ep_web1}
		\frac{d}{d\theta_2} \left[\theta_2 
		\left(E[\min\{p_1,p_2\}] + \left(2 - \frac{2}{N}\right) 
		(E[p]-E [\min\{p_1,p_2\}])\right)\right]<0.
	\end{equation}
	Now, using \eqref{eq:dtau_web}, we rewrite 
	\eqref{eq:Ep_web1} as
	\begin{equation*}
		\frac{d \theta_2 E[\min\{p_1,p_2\}]}{d\theta_2} <0,
	\end{equation*}
	so
	\begin{equation*}
		\theta_2 \left(E[\min\{p_1,p_2\}] + \frac{\partial 
			E[\min\{p_1,p_2\}]}{\partial \mu(q)} \times \frac{d 
			\mu(q)}{d 
			\theta_2}\right)<0.
	\end{equation*}
	We substitute the value of $d\mu(q)/d\theta_2$ from 
	\eqref{eq:dmu_dt} and simplify to obtain
	\begin{equation*}
		\frac{\partial E[p]}{\partial \mu(q)}E[\min\{p_1,p_2\}] 
		- 
		\frac{\partial E[\min\{p_1,p_2\}]}{\partial 
		\mu(q)}E[p]<0.
	\end{equation*}
	Employing the expression for $E[p]$ and 
	$E[\min\{p_1,p_2\}]$, we obtain
	\begin{equation}\label{eq:outlay_web}
		v\frac{2 \mu(q) \left(\mu(q) (1+\mu(q)) \ln	
			^2\left(1+\frac{1}{\mu(q)}\right)-1\right)}{1+\mu(q)}<0,
	\end{equation}
	which is true only if the terms in the large brackets in the 
	numerator are negative, or
	\begin{equation*}
		\ln^2\left(1+\frac{1}{\mu(q)}\right)<\frac{1}{\mu(q) 
			(1+\mu(q)) },
	\end{equation*}
	or
	\begin{equation*}
		- \frac{1}{\sqrt{\mu(q) (1+\mu(q))}}<\ln\left(1 + 
		\frac{1}{\mu(q)}\right)<\frac{1}{\sqrt{\mu(q) 
		(1+\mu(q))}}.
	\end{equation*}
	The left-hand inequality clearly holds for any $\mu(q)>0$.  
	Regarding the right-hand inequality, as both its sides go to 
	infinity as $\mu(q) \to 0$ and converge to $0$ as $\mu(q) 
	\to \infty$, the inequality holds if the derivative of the 
	LHS is less negative than that of the RHS.  The derivative 
	of the LHS is equal to $- \frac{1}{\mu(q)+\mu(q)^2}$ and 
	that of the RHS is $-\frac{1}{\mu(q)+\mu(q)^2} \times	
	\frac{1+2\mu(q)}{2\sqrt{\mu(q)+\mu(q)^2}}$.  The former is 
	less negative 
	than the latter if 	$1+2\mu(q)<2\sqrt{\mu(q)+\mu(q)^2}$, or
	\begin{equation*}
		1 + 4 \mu(q) + 4 \mu(q)^2 > 4 \mu(q) + 4 \mu(q)^2
	\end{equation*}
	which is certainly true for $\mu(q)>0$.  This establishes 
	that \eqref{eq:outlay_web} holds.  This means that costly 
	buyers 	are better-off with an increase $\theta_2$ at the 
	expense of $\theta_j$ for $j\neq 3$.  
	
	Finally, as consumers search more and at the same time are 
	better-off with an increase in $\theta_2$ at the 
	expense of $\theta_j$ for $j\neq 3$, they must pay lower  
	expected price.  
	
	This completes the proof of the proposition.

\end{singlespace}
	
			\newpage
	
	\bibliographystyle{ecta}

\end{document}